\documentclass[fleqn,usenatbib]{mnras}

\usepackage{newtxtext,newtxmath}
\usepackage{comment}

\usepackage[T1]{fontenc}

\DeclareRobustCommand{\VAN}[3]{#2}
\let\VANthebibliography\thebibliography
\def\thebibliography{\DeclareRobustCommand{\VAN}[3]{##3}\VANthebibliography}

\usepackage{graphicx}	
\usepackage{amsmath}	
\usepackage{physics}
\usepackage{float}
\usepackage{lipsum}
\usepackage{multirow}
\usepackage{xcolor}
\usepackage{orcidlink}
\usepackage{hyperref}

\title[Deep JWST clustering HOD analysis]{Accelerated evolution of galaxy host halo masses during Cosmic Dawn from deep JWST clustering}

\author[Nicolò Dalmasso et al.]
{Nicolò Dalmasso$^{1,2}$\thanks{e-mail: ndalmasso@student.unimelb.edu.au}\orcidlink{0000-0002-1850-4050},
Giovanni Ferrami$^{1,2}$\orcidlink{0000-0002-2012-4612},
Nicha Leethochawalit$^{3}$\orcidlink{0000-0003-4570-3159},
Emanuele M. Ventura$^{1,2}$\orcidlink{0000-0003-3502-4929},
\newauthor Michele Trenti$^{1,2}$\orcidlink{0000-0001-9391-305}
\\
$^{1}$School of Physics, University of Melbourne, Parkville, Vic 3010, Australia\\
$^{2}$ARC Centre of Excellence for All Sky Astrophysics in 3 Dimensions (ASTRO 3D), Australia\\
$^{3}$National Astronomical Research Institute of Thailand (NARIT), Mae Rim, Chiang Mai, 50180, Thailand\\
}

\date{Accepted XXX. Received YYY; in original form ZZZ}

\pubyear{2025}

\begin{document}
\label{firstpage}
\pagerange{\pageref{firstpage}--\pageref{lastpage}}
\maketitle

\begin{abstract}
We present the deepest clustering analysis of early galaxies to date, analyzing $N_{\rm{g}} \simeq 6500$ photometrically-selected Lyman Break Galaxies from JWST's Advanced Deep Extragalactic Survey (JADES) to reveal how galaxies and dark matter evolved during cosmic dawn ($5 \leq z < 11$). Using halo occupation distribution (HOD) modeling of the two-point angular correlation function, we trace the galaxy-halo relationships across the first billion years of cosmic history. Our analysis reveals that galaxies at $z = 10.6$ reside in dark matter halos over an order of magnitude less massive ($M_{\rm{h}} \sim 10^{10.12} M_{\odot}$) than their counterparts at $z = 5.5$ ($M_{\rm{h}} \sim 10^{11.45} M_{\odot}$), while exhibiting correspondingly higher effective bias values ($b_{\rm{g}}^{\rm{eff}} = 8.13^{+0.04}_{-0.02}$ compared to $5.64^{+0.10}_{-0.13}$). 
Correspondingly, the satellite galaxy fraction hints at a declining trend with decreasing redshift, reaching $<1\%$ by $z \sim 5-6$. However, the significant systematic and random uncertainties in the data-model comparison prevent us from drawing robust conclusions on the evolution - if any - of the satellite fraction during the epoch of reionization. These results provide the first view of the coevolution between galaxies and dark matter evolved at redshift $\gtrsim 10$, offering additional and independent constraints on early galaxy formation models tuned to reproducing luminosity function evolution.

\end{abstract}

\begin{keywords}
cosmology: observations – galaxies: evolution – galaxies: formation – galaxies: general – galaxies: high-redshift – large-scale structure of Universe
\end{keywords}


\section{Introduction }\label{sec: introduction}

Observations of unexpectedly luminous, massive galaxies at high-redshifts obtained with the James Webb Space Telescope (JWST, e.g., \citealt{Naidu_2022, Labbe_2023, Arrabal_2023, Boyett_2024, Carniani_2024, Napolitano_2025}) have highlighted some discrepancies with the predictions of the $\Lambda$CDM framework on galaxy formation at early times \citep{Padmanabhan_2023}, or tuning of the baryonic processes (e.g., \citealt{Yung_2022, Ferrara_2023, Mason_2023, Kolchin_2023, Gelli_2024}). 
These results suggest a more rapid assembly of baryons than the $\Lambda$CDM model typically predicts, indicating an early Universe with substantial galaxy masses. 
Wide-field JWST surveys, along with spectroscopic analyses of massive galaxy candidates, are essential for assessing this discrepancy and determining if these high-redshift galaxies align with $\Lambda$CDM expectations.

A fundamental approach for testing cosmological models involves investigating the relation between luminous galaxies and the dark matter halo that hosts them by observing the projected spatial arrangement of galaxies, as summarised by the two-point correlation function (e.g., \citealt{Peebles_1980, Bahcall_1983, Davis_1983}). Such analyses are based on galaxy clustering statistics, which explore the spatial distribution and correlations among galaxies. This, in turn, offers essential information regarding the large-scale structure of the Universe and the processes that drive galaxy formation and evolution \citep{Groth_1977}.

A significant application of clustering analysis centers on Lyman Break Galaxies (LBGs), star-forming galaxies identifiable by their dust-unaffected rest-frame UV spectral energy distribution (SED). 
These galaxies offer insights into the relationship between the properties of galaxies and their host halos, probing essential aspects of galaxy evolution. LBGs exhibit strong clustering patterns, with correlation lengths similar to present-day bright spiral galaxies (e.g., \citealt{Giavalisco_1998, Giavalisco_2001, Porciani_2002, Adelberger_2005}). This clustering indicates that LBGs are biased tracers of the mass density field, as they predominantly reside in massive dark matter halos, which display stronger clustering than less massive halos (e.g., \citealt{Bardeen_1986,Mo_1996,Tormen_1999}).

Previous studies have demonstrated a clear link between halo mass and galaxy properties, highlighted by the correlation between LBG clustering strength and UV luminosity. It has been found in past studies that brighter LBGs exhibit larger correlation lengths (e.g., \citealt{Giavalisco_2001, Ouchi_2004, Adelberger_2005}) implying that brighter galaxies are generally hosted within more massive halos. 

Studying the angular correlation function (ACF), which quantifies the increased probability of finding a galaxy at a given angular separation relative to a random distribution, provides valuable insights into substructure and the halo occupation distribution (HOD). However, obtaining reliable measurements with high signal-to-noise (S/N) ratios across a broad luminosity range is crucial, as most satellite galaxies within dark matter halos are typically less luminous than their central counterparts \citep{Giavalisco_2001, Hamana_2004}. 

Distinct galaxy populations, categorized by luminosity, mass, morphology, and intrinsic characteristics, demonstrate unique clustering patterns influenced by selection biases that favor particular halo masses. More massive halos generally exhibit enhanced clustering compared to their less massive counterparts, resulting in galaxies hosted in massive halos appearing more clustered than those in lower mass halos. Furthermore, biases inherent in clustering methodologies and the construction of the ACF affect clustering measurements therefore large samples of galaxies that cover a broad range of luminosities are essential for precise measurements of clustering segregation. 
It is also crucial to account for survey depth and completeness as disparities in observational coverage across different regions may lead to inflated clustering estimates at greater separations if a uniform distribution of galaxies is assumed as outlined in \citet{Dalmasso_2023}. 
These techniques have been employed to study a simple model of galaxy clustering at high redshift, based on a power-law fit to the ACF, using the first data release of JWST Advanced Deep Extragalactic Survey (JADES) \citep{Dalmasso_2024}.

In this study, we present clustering measurements of LBGs at high redshift ($5 \leq z < 11$), based on Version 2.0 of the JADES data release, which includes approximately twice the number of photometrically identified high-redshift galaxies compared to previous releases. Crucially, this expanded sample enables us to extend HOD modeling to the faint end of the luminosity function ($M_{UV} \leq -17.0$) and to higher redshifts than previously accessible, thereby probing galaxy-halo connections in unexplored regions of parameter space.
We model the observed ACF using a HOD framework to disentangle the contributions from central and satellite galaxies to the clustering signal.
We further investigate the evolution of the average halo occupation to constrain the satellite fraction at cosmic dawn and assess the consistency of these results with $\Lambda$CDM predictions for early galaxy formation.

The structure of this paper is outlined as follows: in Sec.\ref{sec: data sets and sample selection}, we describe data reduction and the creation of the parent sample; in Sec.\ref{sec: clustering analysis}, we present the ACF fits of the HOD model used in this study; in Sec.\ref{sec: DMH clustering estimations}, we discuss our results; and in Sec.\ref{sec: summary}, we offer a summary of our key findings. 
We adopt the cosmological parameters determined by the Planck Collaboration 
\citep{Ade_Planck_2015}: $(\omega_{\rm{m}}, \Omega_{\lambda}, h, \sigma_{8}) = (0.3075, 0.6925, 0.6774, 0.8159)$. 
Magnitudes are reported in the AB system \citep{Oke_1983}.

\section{Data sets and sample selection}\label{sec: data sets and sample selection}

\subsection{JADES GOODS-South Data}\label{subsec: data points catalog}

In this study, we analyze imaging data from the public release targeting the Great Observatories Origins Deep Survey-South (GOODS-South) field, provided by the JADES\footnote{\url{https://archive.stsci.edu/hlsp/jades}\label{note: jades}} collaboration \citep{Bunker_2023,Eisenstein_2023a,Eisenstein_2023b,Hainline_2023,Rieke_2023, D'Eugenio_2024}. This data set, captured by the NIRCam instrument, spans 67.7 arcminutes$^2$ across the Deep and Medium programs, utilizing nine filters (F090W, F115W, F150W, F200W, F277W, F335M, F356W, F410M, F444W) covering a wavelength range of $0.8 - 5.0 \mu m$ with a spatial resolution of 0.03 arcsec/pixel. A detailed description of the JADES Version 2.0 data release, which includes the galaxy candidates studied here, can be found in \citet{Eisenstein_2023b}.

Our analysis focuses on galaxies within a redshift range of $5 \leq  z < 11$, divided into six bins with a width of $\Delta z = 1.0$ with the resulting average redshifts $\overline{z}=$ 5.5, 6.5, 7.4, 8.5, 9.3 and 10.6. We select candidates based on a signal-to-noise ratio (SNR) of at least 5.0 in the F200W NIRCam band, ensuring that they are confined within observable regions on both root mean square (rms) and segmentation maps to improve the accuracy of our results.

To reduce uncertainties caused by varying depths and multiple exposures in different regions, which could impact clustering estimates, we estimate the completeness of the survey ($\mathscr{C}$). As in \citet{Dalmasso_2024}, we focus on the faintest observable galaxies and apply a magnitude threshold of $M_{UV} < -17.0$ for all cataloged candidates.

Detection completeness was measured using the injection-recovery tool \texttt{GLACiAR2} \citep{Leethochawalit_2022} as a function of magnitude and redshift. Completeness decreases from $\sim$80\% at bright magnitudes ($M_{\mathrm{UV}} < -18$) to $\sim$50\% around $M_{\mathrm{UV}} \sim -15.5$ to $-16.5$ at intermediate redshifts ($z \sim 5$--$9$), with a notable shift to brighter absolute magnitudes at $z=10.5$. The 50$\%$ completeness limits for each redshift bin are presented in Tab.\ref{tab:completeness}.

\begin{table}
\centering
\begin{tabular}{ccc}
\hline\hline
z & $M_{\mathrm{UV},50\%}$ & $m_{\mathrm{AB},50\%}$ \\
\hline
5.5  & $-15.50$ & 31.10 \\
6.5  & $-15.50$ & 31.37 \\
7.5  & $-15.50$ & 31.59 \\
8.5  & $-15.50$ & 31.78 \\
9.5  & $-16.50$ & 30.95 \\
10.5 & $-19.67$ & 27.93 \\
\hline
\end{tabular}
\caption{Magnitude limits of our galaxy samples, defined as the value where injection-recovery simulations achieve 50$\%$ source detection efficiency. Absolute magnitudes are in the rest-frame UV; apparent magnitudes are in F444W.}
\label{tab:completeness}
\end{table}
\subsection{Random points catalog}\label{subsec: random points catalog}

To perform clustering analysis, it is essential to create a complementary catalog of simulated galaxies. These simulated galaxies play a key role in defining the ACF estimator, particularly in the calculation of RR($\theta$), the pair count of random points. These random points are generated to mimic the observational conditions, reflecting any selection biases in the parent sample.

Due to variations in observational depth, survey completeness is not uniform across the field. As a result, certain portions of the imaging field may exhibit an apparent excess or deficit of sources, potentially influenced by angular patterns correlated with the rms maps. To correct for these effects, we perform an artificial source injection and recovery process, generating random points within the ACF measurement area to model the completeness variation.

The random point catalog generation follows the recovery procedure detailed in \citet{Dalmasso_2023}. Briefly, we used the injection-recovery tool \texttt{GLACiAR2} \citep{Leethochawalit_2022} to inject galaxies into JADES images. These galaxies were distributed across redshift bins ranging from $z_{\rm min}=4.5$ to $z_{\rm max}=13.0$ in steps of 0.5, and UV magnitudes from $M_{\rm{min}}=-25$ to $M_{\rm max}=-13$ in 0.5 mag decrements. For each redshift-magnitude bin, we injected $N=3200$ galaxies at random positions, corresponding to approximately one galaxy per 30 arcsec$^2$. The injected galaxies were modeled with disk-like light profiles (Sersic index $n=1$), random inclinations, and ellipticities. The spectral energy distributions (SEDs) were drawn randomly from the JAGUAR mock catalog \citep{Williams2018}, matching the redshift bin. Galaxy recovery followed the same process used for the real data \citep{Rieke_2023}, utilizing a detection image constructed from the F227W, F335M, F356W, F410M, and F444W bands.

To finalize the random points catalog, we selected recovered candidates and employed a Monte Carlo hit-and-miss method. Each simulated galaxy in the random catalog was assigned a probability of detection, defined by:
\begin{equation}
\text{p}(M) = \frac{\Phi(M)}{\Phi(M_{\text{lim}})},
\end{equation}
where $M$ is the galaxy's absolute magnitude, and $M_{\text{lim}}$ is the limiting magnitude set at $M_{UV}=-17.0$. The luminosity function parameters evolve with redshift, following the Schechter profile described by \citet{Bouwens_2021}.

\section{Clustering Analysis }\label{sec: clustering analysis}

\subsection{Angular Correlation Function (ACF) estimation }\label{subsec: ACF}

The ACF is characterized by the observable $\omega_{\text{obs}}(\theta)$ as defined by \citet{Landy_1993}:
\begin{equation}\label{eq: estimator}
    \omega_{\text{obs}}(\theta)= \frac{\text{DD}(\theta) - 2\text{DR}(\theta) + \text{RR}(\theta)}{\text{RR}(\theta)},
\end{equation}
which measures the excess probability of finding pairs of objects in the parent sample at a given angular separation compared to a random distribution of candidates within the same survey area.

In Eq.\ref{eq: estimator} the term $\text{DD}(\theta)$ represents the number of pairs of galaxies within an angular separation range of $(\theta \pm \delta\theta)$ (data-data pairs), $\text{DR}(\theta)$ denotes pairs formed by one observed galaxy and one randomly generated galaxy (data-random pairs), and $\text{RR}(\theta)$ corresponds to the pair count from the randomly generated catalog (random-random pairs). In this study, we consider an angular range of $\theta \in (0.2,850)$ arcsec, using a variable number of equally spaced angular bins.

The ACF, as defined above, is typically underestimated due to the finite size of the observational survey. To correct for this bias, we introduce a coefficient known as the integral constraint (IC) \citep{Groth_1977,Peacock_1991}:
\begin{equation}\label{eq: IC}
    \text{IC} =\frac{\sum_{i} \text{RR}(\theta_{i}) \omega_\text{model}(\theta)}{\sum_{i} \text{RR}(\theta_{i})},
\end{equation}
where $\omega_\text{model}(\theta)$ is the best-fit model ACF and $\text{RR}(\theta_i)$ is the pair count of randomly generated galaxies within a specific angular bin. This coefficient is used to correct the observed ACF, linking the observed and true ACF measurements as follows:
\begin{equation}\label{eq: w_true}
    \omega_\text{true}(\theta)=\omega_\text{obs}(\theta) + \text{IC}.
\end{equation}

To accurately account for this coefficient in the ACF estimation, we considered $\text{RR}(\theta_i)$ up to the maximum angular separation covered by our survey area. This ensures that, at large angular separations, the IC approaches zero preventing any bias in our measurements. Notably, both the IC and $\omega_\text{model}(\theta)$ are determined simultaneously during the model fitting process, as outlined in Sec.\ref{subsec: HOD model}.

To quantify the uncertainties in the ACF measurements, we construct the normalized covariance matrix using the standard estimator:
\begin{equation}\label{eq: cov}
C_{ij} = \frac{1}{N_{\text{boot}} - 1} \sum_{l=1}^{N_{\text{boot}}} \left[ \omega^l (\theta_i) - \overline{\omega}(\theta_i) \right] \left[ \omega^l (\theta_j) - \overline{\omega}(\theta_j) \right],
\end{equation}
where $N_{\text{boot}}$ is the total number of resamplings performed, $\omega^l (\theta_i) $ is the cross-correlation function measured from each realization in the $i$-th bin, and $\overline{\omega}(\theta_i) $ is the mean of the cross-correlation function in the same bin. The uncertainty associated with each bin is taken to be the square root of the corresponding diagonal element in the covariance matrix.

\subsection{Halo Occupation Distribution (HOD) model}\label{subsec: HOD model}

To interpret galaxy clustering, we apply the HOD formalism, which describes how galaxies populate dark matter halos
(\citealt{HOD_Ma_Fry_2000, HOD_Peacock_Smith_2000, HOD_Seljak_2000}).
The underlying assumption of the model is that all dark matter halos are spherical with a density distribution that depends only on their mass.
In addition, we assume that the average number of galaxies residing in each halo depends on the halo mass\footnote{The outcome of these assumptions on galaxy clustering have been well tested on both numerical simulations and observations in the local/low redshift universe.}. 

The galaxy population is divided between central and satellite galaxies. 
Central galaxies reside at the center of the dark matter halo, while satellites reside at the center of a sub-halo and orbit around central galaxies inside a larger host halo.
The mean number of galaxies $\langle N \rangle$ residing in a halo of mass $M_{\rm{h}}$ is then the sum of the average of central and satellite galaxies, 
\begin{equation}\label{eq:N_total}
    \langle N  (M_{\rm{h}}) \rangle= \langle N_{\rm{c}} (M_{\rm{h}})\rangle + \langle N_{\rm{s}}  (M_{\rm{h}})\rangle.
\end{equation}
We adopt expressions of the number of central galaxies $\langle N_{\rm{c}} \rangle$ and the number of satellite galaxies $\langle N_{\rm{s}} \rangle$ motivated by N-body and smoothed particle hydrodynamics simulations (e.g., \citealt{Kravtsov_2004, Zheng_2005, Garel_2015}) and defined as
\begin{equation}\label{eq:N_central}
    \langle N_{\rm{c}} (M_{\rm{h}}) \rangle = 
    \frac{1}{2}\left[ 1 + 
    \rm{erf}\left( 
    \frac{\log M_{\rm{h}} - \log M_{\rm{min}}}{\sqrt{2}\sigma_{\log M_{\rm{h}}}}
    \right) \right]
\end{equation}
and
\begin{equation}\label{eq:N_satellite}
    \langle N_{\rm{s}}  (M_{\rm{h}}) \rangle = 
    \langle N_{\rm{c}} (M_{\rm{h}}) \rangle
    \left( 
    \frac{M_{\rm{h}}}{M_{\rm{sat}}}
    \right)^\alpha \:,
\end{equation}
respectively.
The values of $M_{\rm{min}}$ and $\sigma_{\log M_{\rm{h}}}$ define the inflection point and width of the sigmoid function describing $\langle N_{\rm{c}} \rangle$.
The values of $M_{\rm{sat}}$ and $\alpha$ represent the amplitude and slope of a power law factor that determines the average number of satellites.
We fix $\sigma_{\log M_{\rm{h}}} = 0.2$ following previous studies (e.g., \citealt{Kravtsov_2004, Zheng_2005, Conroy_2006, Harikane_2016}).

In this paper, we are interested in testing whether the angular clustering resulting from the HOD formalism is compatible with the distribution of galaxies over a large range of luminosities at  $z \gtrsim5$, and comparing the resulting halo-luminosity relation to theoretical predictions of the $\Lambda$CDM paradigm. 
In order to do so, we need to relate the average number of galaxies, determined by the values of $M_{\rm{min}}$, $M_{\rm{sat}}$ and $\alpha$, to the measured ACF.
This is achieved by estimating the ACF at a angular distance $\theta$ from the galaxy power spectrum $P_{\rm{g}}$ via the Limber approximation \citep{Bartelmann_2001} projected over a the normalized redshift distribution $\mathcal{N}(z)$ of the observed sample 
\begin{equation}\label{eq:ACF}
    \omega (\theta) = 
    \int \dd z \mathcal{N}(z)\dv{z}{r} 
    \int \dd k \frac{k}{2\pi} 
    P_{\rm{g}}(k, z) J_0[\theta r(z) k] \:,
\end{equation}
where $r(z)$ is the radial comoving distance and $J_0$ is the zeroth-order Bessel function of the first kind.

The galaxy power spectrum receive contributions by the 1-halo term and 2-halo term expressed as:
\begin{equation}\label{eq:Pg}
    P_{\rm{g}} (k, z) = 
    P_{\rm{g}}^{\rm{1h}} (k, z) + 
    P_{\rm{g}}^{\rm{2h}} (k, z).
\end{equation}

The contribution to the power spectrum coming from pairs of galaxies contained within the same halo is
\begin{equation}\label{eq:P1h}
    P_{\rm{g}}^{\rm{1h}} (k, z) = 
    P_{\rm{g}}^{\rm{cs}} (k, z) + 
    P_{\rm{g}}^{\rm{ss}} (k, z) \:,
\end{equation}
and is split into the contributions of the central-satellite and satellite-satellite galaxies pairs, defined as 
\begin{equation}\label{eq:Pcs}
    P_{\rm{g}}^{\rm{cs}}(k, z) = 
    \frac{2}{n_{\rm{g}}^{2}}
    \int \dd M_{\rm{h}} \langle N_{\rm{c}}N_{\rm{s}} (M_{\rm{h}})\rangle
    \dv{n}{M_{\rm{h}}} (M_{\rm{h}}, z) u(k, M_{\rm{h}}, z)
\end{equation}
and
\begin{equation}\label{eq:Pss}
    P_{\rm{g}}^{\rm{ss}}(k, z) = 
    \frac{1}{n_{\rm{g}}^{2}}
    \int \dd M_{\rm{h}} \langle N_{\rm{s}}(N_{\rm{s}}-1) (M_{\rm{h}})\rangle
    \dv{n}{M_{\rm{h}}} (M_{\rm{h}}, z) u^2(k, M_{\rm{h}}, z) \: ,
\end{equation}
respectively. 
On the other hand, the contributions to the galaxy power spectrum introduced by pairs residing in different halos is
\begin{equation}\label{eq:P2h}
    P_{\rm{g}}^{\rm{2h}} (k, z) = 
    \left[
    \frac{1}{n_{\rm{g}}}
    \int \dd M_{\rm{h}} \langle N(M_{\rm{h}})\rangle
    \dv{n}{M_{\rm{h}}} (M_{\rm{h}}, z) b_{\rm{h}}(M_{\rm{h}}, z) u(k, M_{\rm{h}}, z)
    \right]^2 \: ,
\end{equation}
where $b_{\rm{h}}$ the halo bias factor from \citep{Tinker_2010} in the linear regime.
Here, $u(k, M_{\rm{h}}, z)$ is the Fourier transform of the dark matter halo NFW density profile normalized by its mass (e.g., \citealt{Cooray_2002}), adopting the concentration parameter relation from \citep{Correa_2015}.
Finally, $n_{\rm{g}}(z)$ is the galaxy number density obtained weighting the halo mass function from \citet{Behroozi_2010} on the average occupation number  
\begin{equation}\label{eq:ngz}
    n_{\rm{g}} (z) = 
    \int \dd M_{\rm{h}} \dv{n}{M_{\rm{h}}} (M_{\rm{h}}, z)
    \langle N(M_{\rm{h}})\rangle \: ,
\end{equation}
where the halo mass function $\dv{n}{M_{\rm{h}}}$ provides the comoving number density of halos as a function of halo mass and redshift, calibrated using N-body simulations and validated at high redshifts (\citealt{Tinker_2008, Behroozi_2010}). The quantity $\bar{n}_{\rm{g}}$ is the integral of $n_{\rm{g}}(z)$ over comoving volume averaged over the redshift distribution of the sample
\begin{equation}\label{eq:avg_ng}
    \overline{n}_{\rm{g}} = \frac{
    \int \dd z \dv{V(r)}{z} \mathcal{N}(z) n_{\rm{g}}(z)}
    {\int \dd z \dv{V(r)}{z} \mathcal{N}(z)} \: .
\end{equation}

To analyze the relation between the distribution of galaxies within their host halos, we also define the following average quantities: the average satellite fraction
\begin{equation}\label{eq:fsat}
    f_{\rm{sat}} (z)= 
    \frac{1}{\overline{n}_{\rm{g}}} 
    \int \dd M_{\rm{h}} \dv{n}{M_{\rm{h}}} (M_{\rm{h}}, z)
    \langle N_{\rm{s}}(M_{\rm{h}})\rangle\: ,
\end{equation}
the effective bias
\begin{equation}\label{eq:beff}
    b^{\rm{eff}}_{\rm{g}} (z)= 
    \frac{1}{\overline{n}_{\rm{g}}} 
    \int \dd M_{\rm{h}} \dv{n}{M_{\rm{h}}} (M_{\rm{h}}, z)
    \langle N(M_{\rm{h}})\rangle  b_{\rm{h}}(M_{\rm{h}}, z) \: ,
\end{equation}
and the average halo mass
\begin{equation}\label{eq:avg_Mh}
    \langle M_{\rm{h}}\rangle (z)= 
    \frac{1}{\overline{n}_{\rm{g}}} 
    \int \dd M_{\rm{h}} \dv{n}{M_{\rm{h}}} (M_{\rm{h}}, z) M_{\rm{h}}
    \langle N(M_{\rm{h}})\rangle \: .
\end{equation}

We expect these average quantities to be weakly dependent on the approximation of a linear evolution of the halo bias factor adopted in our HOD model.
In fact, while the one-halo term, and therefore the satellite fraction, is primarily determined by the ACF below 5 arcsec and the two-halo term is well constrained by the observed ACF between $10^2$ and $10^3$ arcsec,
the non-linear bias has only a percent-level effect on angular clustering at $z\ll1$\citep{Bosch_2013}, and contributes to two-halo term primarily around $\theta \sim 10$ arcsec at $3 < z < 5$ (\citealt{Jose_2016,Jose_2017}). 
Nonetheless, the impact of the linear bias approximation on galaxy clustering remains untested at the high redshifts and low luminosities probed in this study.
We defer to future work an in-depth comparison between our results and dedicated simulations of high-$z$ galaxies that can fully account for non-linear bias effects.

\section{Dark Matter Halo Clustering estimations}\label{sec: DMH clustering estimations}

\subsection{HOD model fitting}\label{subsec: ACF-HOD fit}

\begin{figure*}
  \includegraphics[width=1\linewidth]{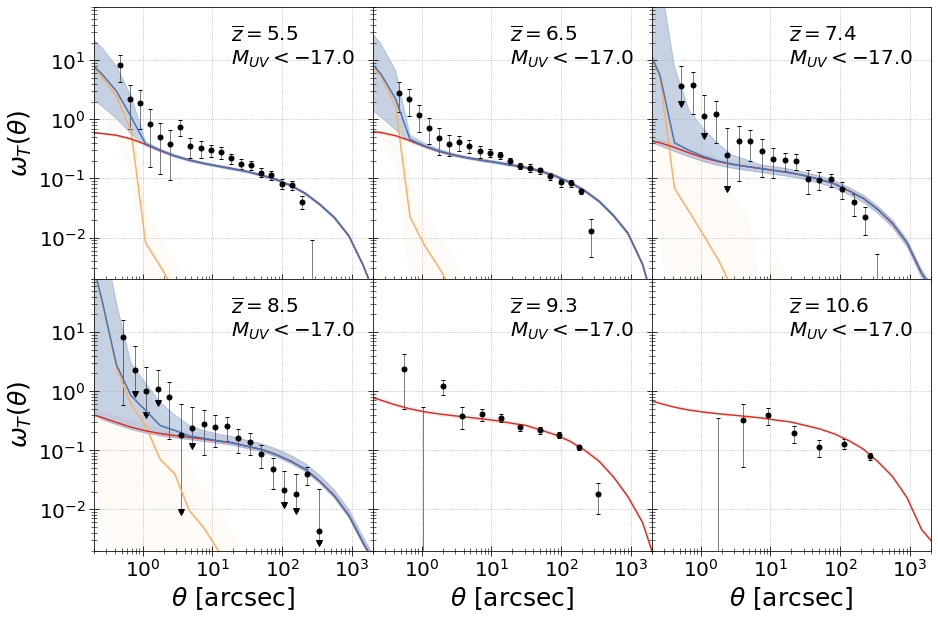}
\caption{Measured ACFs in the GOODS-S field at six mean redshifts from bootstrap resampling with $1\sigma$ uncertainties derived from the covariance matrix. Blue line represent the HOD best-fit model ($\omega_{\rm{mod}}(\theta)$), with orange and red lines showing the one-halo ($\omega_{\rm{1h}}$) and two-halo ($\omega_{\rm{2h}}$) terms. Shaded areas indicate $1\sigma$ uncertainties. Mean redshift and absolute UV magnitude thresholds (F200W band) are shown in the upper right of each panel.}
\label{fig: ACF-HOD-COV}
\end{figure*}

\begin{figure}
  \includegraphics[width=1\linewidth]{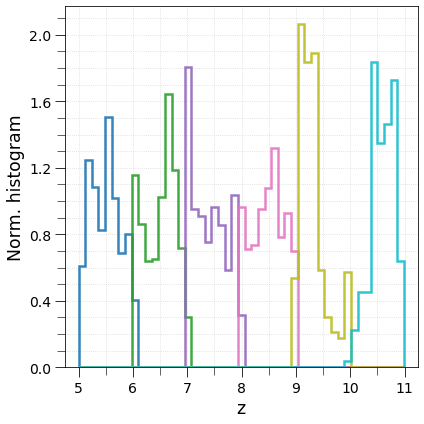}
  \caption{Normalized redshift distributions of the parent galaxy samples, with fine structure resulting from a combination of photometric redshift selection efficiency and sample variance in excess of Poisson noise due to clustering.}
\label{fig: nz_hist}
\end{figure}

\begin{table*}
\centering
\begin{tabular}{cccccccc}
    \hline \hline
    \multicolumn{8}{c}{HOD clustering measurements}\\
    \hline
    $\overline{z}$ & $N_{\rm{g}}$ & $<M_{UV}>$ & $\log(M_{\rm{min}}/M_\odot)$ & $\log(M_{\rm{sat}}/M_\odot)$ & $\log(f_{\rm{sat}})$ & $b_{\rm{g}}^{\rm{eff}}$ & $\log(M_{\rm{h}}/M_\odot)$\\
    (1) & (2) & (3) & (4) & (5) & (6) & (7) & (8)\\ 
    \hline \hline 
    $5.5$  & $1686$ & -17.90 & $11.57^{+0.06}_{-0.08}$ & $14.58^{+0.52}_{-0.36}$ & $-3.13^{+0.32}_{-0.49}$ & $5.64^{+0.10}_{-0.13}$ & $11.45^{+0.02}_{-0.04}$\\[.2cm]
    $6.5$  & $2937$ & -17.87 & $11.16^{+0.05}_{-0.05}$ & $14.18^{+0.86}_{-0.42}$ & $-2.99^{+0.38}_{-0.82}$ & $6.41^{+0.15}_{-0.15}$ & $11.18^{+0.03}_{-0.04}$\\[.2cm]
    $7.4$  & $909$ & -17.77 & $10.61^{+0.14}_{-0.19}$ & $12.82^{+2.14}_{-1.39}$ & $-2.17^{+1.23}_{-2.05}$ & $5.92^{+0.31}_{-0.36}$ & $10.65^{+0.08}_{-0.10}$\\[.2cm]
    $8.5$  & $712$ & -17.88 & $10.05^{+0.24}_{-0.30}$ & $10.72^{+3.42}_{-1.95}$ & $-0.54^{+0.52}_{-3.17}$ & $6.20^{+0.18}_{-0.15}$ & $10.37^{+0.05}_{-0.02}$\\[.2cm]
    $9.3$  & $459$ & -17.82 & $8.95^{+0.64}_{-0.65}$ & $-$ & $-$ & $6.95^{+0.13}_{-0.08}$ & $10.23^{+0.02}_{-0.04}$\\[.2cm]
    $10.6$ & $218$ & -17.81 & $8.82^{+0.61}_{-0.56}$  & $-$ & $-$ & $8.13^{+0.04}_{-0.02}$ & $10.12^{+0.08}_{-0.06}$\\

    \hline \hline
\end{tabular}
\caption{Summary of the clustering results from the HOD model best fit. Columns: (1) Mean redshift, (2) Number of galaxies, (3) Average UV absolute magnitude of the sample in the F200W band , (4) Best-fit value of $M_{\rm{min}}$, (5) Best-fit value of $M_{\rm{sat}}$, (6) Fraction of galaxy satellites in the dark matter halo (Eq.\ref{eq:fsat}), (7) Effective galaxy bias (Eq.\ref{eq:beff}), (8) Mean dark matter halo mass $M_{\rm{h}}$ (Eq.\ref{eq:avg_Mh}).}
\label{tab: HOD results}
\end{table*}

To fit the ACF results with our HOD model, we employ a Markov Chain Monte Carlo (MCMC) method, allowing two parameters ($M_{\rm{min}}$, $M_{\rm{sat}}$) to vary freely during the process. Following findings from previous works we decided to fix $\alpha=1$ for the analysis (i.e., \citealt{Kravtsov_2004,Zheng_2005, Conroy_2006, Zehavi_2011, Harikane_2016, Ishikawa_2017, Harikane_2021}). 

The likelihood function adopted is defined as:
\begin{equation}\label{eq: likelihood cov}
    \ln(\mathscr{L}) = - \frac{1}{2} \sum_{i,j}\left[ \omega(\theta_i) - \omega_{\rm{m}}(\theta_i) \right] C_{ij}^{-1} \left[ \omega(\theta_j) - \omega_{\rm{m}}(\theta_j) \right],
\end{equation}
where $\omega(\theta)$ represents the corrected ACF measurements from Eq.\ref{eq: w_true}, $\omega_{\rm{m}}(\theta)$ are the model predictions from the HOD in Eq.\ref{eq:ACF}, and $C_{ij}^{-1}$ is the inverse covariance matrix from Eq.\ref{eq: cov}.

For the MCMC fitting, we applied flat priors of $\log(M_{\rm{min}}/M_{\odot}), \log(M_{\rm{sat}}/M_{\odot}) \in [8, 16]$ for both free parameters. We ran 32 walkers for up to 2000 steps, starting from literature-based initial values (i.e., \citealt{Zheng_2005, Harikane_2016,Harikane_2021, Shuntov_2022, Shuntov_2025,Paquereau_2025}), with convergence determined by $30\tau<N_{\rm iter}$ and $\Delta\tau/\tau<15\%$ \citep{Paquereau_2025}.

In Fig.\ref{fig: ACF-HOD-COV}, we present a comparison between the measured ACFs of LBGs candidates in the GOODS-S field from JADES data release Version 2.0 and the ACFs predicted using our HOD formalism. The blue line represents our best-fit model based on the 50th percentile of the MCMC-derived parameters, while the shaded region encompasses the model predictions for the 16th and 84th percentile ranges of ($M_{\rm{min}}$, $M_{\rm{sat}}$). The one-halo term $\omega_{1h}(\theta)$ and two-halo term $\omega_{2h}(\theta)$ are represented by yellow and red solid lines with their associated uncertainty regions, respectively. Tab.\ref{tab: HOD results} summarizes the best-fit HOD parameters from our MCMC analysis along with three key clustering metrics: the satellite fraction $f_{\rm{sat}}$, effective galaxy bias $b_{\rm{g}}^{\rm{eff}}$, and dark matter halo mass $M_{\rm{h}}[M_{\odot}]$. 

In all six panels of Fig.\ref{fig: ACF-HOD-COV} we observe a non-optimal fitting of the correlation function by the MCMC analysis with the implemented HOD model. These systematic discrepancies show under-prediction at intermediate scales ($\theta \sim 3-10$ arcsec) and over-prediction at large scales ($\theta \gtrsim 100$ arcsec). We attribute the data-modeling mismatch to a combination of intrinsic limitations of the HOD modeling and of specific assumptions in our approach, such as fixing a subset of the free parameters due to the limited number of datapoints available at these high redshifts. Furthermore, an element that may impact the fit quality is the adoption of a linear evolution for the halo bias factor, following \cite{Tinker_2010}. This linear approximation is thought to be well justified in the low to intermediate redshift regime $z<5$ for relatively bright sources. However, its impact on clustering studies during the epoch of reionization, and for the low-luminosity (sub $L_*$) galaxies observed by JWST remains untested. Therefore this may contribute significantly to the observed model-data discrepancies and will warrant detailed further investigations. Importantly, the quality of the fit in Fig.\ref{fig: ACF-HOD-COV} is not unique to our analysis but represents a common issue in HOD modeling of high-redshift galaxy clustering. In fact, similar systematic deviations between model predictions and observations are evident in multiple studies at these epochs including \citealt{Harikane_2016, Harikane_2021}, \cite{Paquereau_2025} and \cite{Shuntov_2025}. These examples demonstrate that limited data points and inherent model assumptions produce comparable fitting difficulties across independent analyses of high-$z$ galaxies.

Furthermore, for the highest redshift samples considered here ($z>9$, last two panels in Fig.\ref{fig: ACF-HOD-COV}), there were not enough galaxies in the sample to measure the two point correlation function at small angular separations. Therefore, we performed the fit using only the two-halo term $\omega_{2h}(\theta)$, constraining solely $M_{\rm{min}}$.

Our analysis utilizes a sample of LBGs spanning a wide redshift range during cosmic dawn ($5 \leq  z < 11$) that probes the faint end of the luminosity function from $M_{\rm{UV}}<-17.0$ at various redshift cf. normalized redshift distribution in Fig.\ref{fig: nz_hist}). While direct one-to-one comparisons with existing literature are challenging due to our unique sample characteristics, we can identify general evolutionary trends in the HOD parameters.

As shown in Tab.\ref{tab: HOD results} the free parameters in our HOD model evolve with redshift within our measurement uncertainties. Specifically our findings align with established trends in the literature, we observe $M_{\rm{min}}$ decreasing with increasing redshift, consistent with previous studies by \citet{Hamana_2004}, \citet{Conroy_2006}, and \citet{Harikane_2016, Harikane_2021}. Similarly, $M_{\rm{sat}}$ shows a decreasing trend with redshift, in agreement with \citet{Hamana_2004} and \citet{Conroy_2006}.

\subsection{Satellite Fraction, Effective Bias and Dark Matter Halo Mass}\label{subsec: clustering params}

\begin{figure}
  \includegraphics[width=1\linewidth]{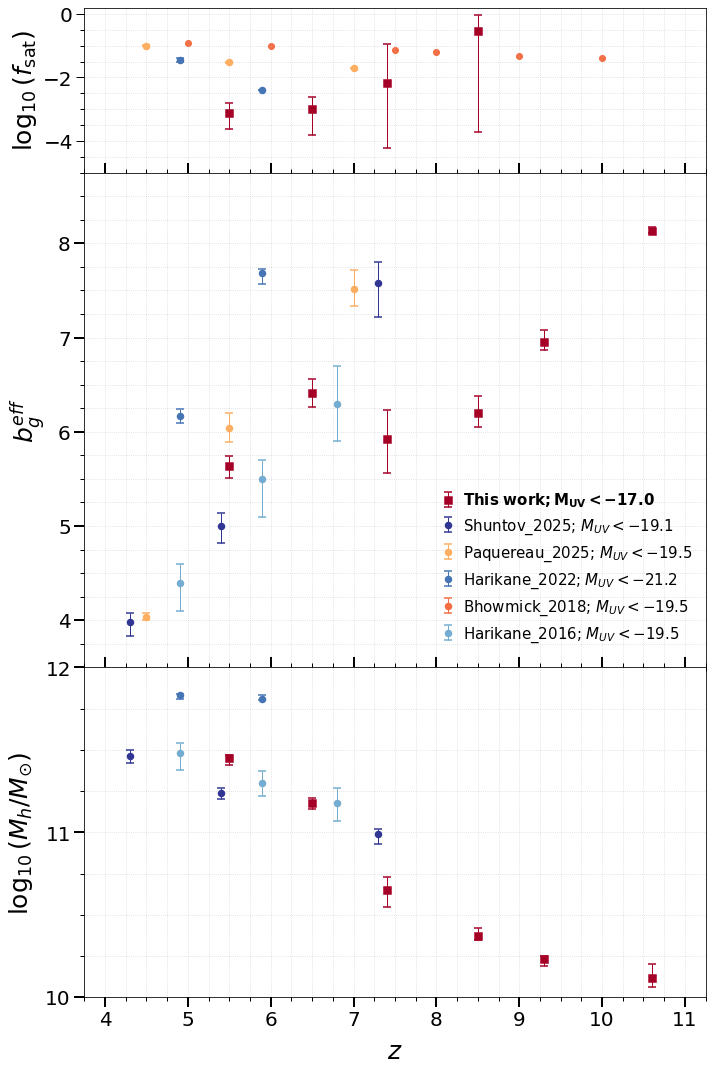}
  \caption{Evolution of clustering parameters as a function of redshift: satellite fraction $f_{\rm{sat}}$ (top panel), effective galaxy bias $b_{\rm{g}}^{\rm{eff}}$ (middle panel), and characteristic halo mass $M_{\rm{h}}[M_{\odot}]$ (bottom panel). Results from this work are shown as solid red squares, while previous HOD modelling studies are color-coded according to the legend (\citealt{Harikane_2016,Bhowmick_2018,Harikane_2021,Paquereau_2025,Shuntov_2025}).}
\label{fig: fsat_beff_Mh_vs_z}
\end{figure}

Fig.\ref{fig: fsat_beff_Mh_vs_z} presents a comparison of the clustering parameters $f_{\rm{sat}}$, $b_{\rm{g}}^{\rm{eff}}$, and $M_{\rm{h}}[M_{\odot}]$ derived from our HOD estimations with results from previous studies employing HOD modelling (\citealt{Harikane_2016,Bhowmick_2018,Harikane_2021,Paquereau_2025,Shuntov_2025}).

The effective galaxy bias measurements demonstrate a clear increasing trend with redshift, rising from $b_{\rm{g}}^{\rm{eff}}=5.64_{-0.13}^{+0.10}$ at $z = 5.5$ to $b_{\rm{g}}^{\rm{eff}}=8.13_{-0.02}^{+0.04}$ at $z = 10.6$. We note an apparent local dip at $z=7.4$ where $b_{\rm{g}}^{\rm{eff}}=5.92_{-0.36}^{+0.31}$, but this measurement is consistent with the neighboring redshift bins within $\sim1.5-2\sigma$, likely reflecting statistical scatter due to cosmic variance and the systematic uncertainties in HOD modeling discussed above. The overall increasing trend is in agreement with theoretical predictions from hierarchical structure formation models \citep{Mo_1996, Tormen_1999}, which predict that galaxies of a given luminosity should reside in increasingly biased (rarer) regions of the cosmic density field at higher redshifts.

Our measurements are broadly consistent with previous observational studies employing HOD modeling \citep{Harikane_2016,Harikane_2021,Paquereau_2025,Shuntov_2025}, though we note that our bias values for galaxies with $M_{\rm{UV}} < -17.0$ appear somewhat higher than those reported for brighter samples at similar redshifts. For instance, \cite{Harikane_2016} and \cite{Shuntov_2025} find lower bias for galaxies with $M_{\rm{UV}} < -19.5$ and $M_{\rm{UV}} < -19.1$, respectively. However, multiple factors complicate direct comparison. First, while \cite{Harikane_2016} reports clustering measurements up to $z \sim 7$, overlapping with the lowest redshift bins of our work, their strongest constraints are obtained at $z = 4$–6. Since galaxy bias changes significantly with redshift and luminosity at $z>5$, even small offsets in the effective redshift and/or luminosity distribution of the samples can lead to appreciable differences in the derived bias values.Additionally, at the faint end of the luminosity function probed by our sample, the relationship between UV luminosity and halo mass may differ from expectations based on brighter galaxies, particularly if star formation efficiency or dust obscuration varies systematically with halo mass at high redshift. A comprehensive understanding of the luminosity-halo mass relation across the full range of UV magnitudes at $z > 6$ will require larger survey volumes and joint constraints from clustering and abundance matching, confirming that high-redshift galaxy populations trace progressively more biased environments as we probe earlier cosmic epochs.

For completeness, we also applied a standard power-law fit to the ACFs analyzed in this work, using the same methodology adopted in previous non-HOD clustering studies. This analysis yields lower effective bias values, ranging from $b_g = 3.15 \pm 0.21$ at $z=5.5$ to $b_g = 8.14 \pm 1.13$ at $z=9.3$, consistent with earlier power-law–based measurements (i.e. \citealt{Overzier_2006, Barone_2014, Harikane_2016, Qiu_2018, Dalmasso_2023, Dalmasso_2024}). Therefore, the offset between the power-law–derived galaxy bias and the HOD-derived effective galaxy bias reflects the known systematic differences between the two estimators rather than any discrepancy in the underlying data.

Our analysis also shows a significant decrease of the characteristic dark matter halo mass $M_{\rm{h}}[M_{\odot}]$ with redshift. The halo mass decreases by more than one order of magnitude from $\log_{10}(M_{\rm{h}}/M_{\odot}) = 11.45_{-0.04}^{+0.02}$ during the final stages of cosmic reionization at $z = 5.5$ to $\log_{10}(M_{\rm{h}}/M_{\odot}) = 10.12_{-0.06}^{+0.08}$ at $z = 10.6$, during cosmic dawn.

In this scenario, galaxies at high redshift reside in progressively less massive dark matter halos compared to similar galaxy populations at lower redshift. Correspondingly, the light-to-mass ratio (relative to the dark matter halo mass) increases by approximately an order of magnitude between $z \sim 6$ and $z \sim 10$ considering that our sample has a consistent selection based on a luminosity-limit to $M_{UV}<-17.0$ and that the average luminosity has limited evolution across redshift. This clearly indicates enhanced star formation efficiency, as the alternative of reduced halo occupation at earlier cosmic epochs is unlikely to apply.

While this study does not aim to provide a physical explanation for this evolution, it offers robust clustering-based measurements that are independent of luminosity function analyses, complementing the trends reported by \citet{Hamana_2004}, \citet{Conroy_2006}, \citet{Harikane_2016, Harikane_2021} and \citet{Shuntov_2025}. These trends are consistent with hierarchical structure formation, where the galaxy-halo relationship evolves as cosmic structure assembles over time.

The redshift evolution of the satellite galaxy fraction $f_{\rm{sat}}$ (top panel of Fig.\ref{fig: fsat_beff_Mh_vs_z}) reveals changes in galaxy clustering properties throughout the early Universe. At the final stages of cosmic reionization ($z \sim 5-6$), our HOD modeling indicates that satellite galaxies constitute less than $1\%$ of the total galaxy population across UV magnitudes down to $M_{\rm UV}\simeq-17.0$.

At the highest redshift probed in our sample ($z\sim7-8$), our measurements suggest an enhancement in satellite populations, with $f_{\rm sat}$ reaching values around $\sim28\%$, though we caution that these measurements carry substantial uncertainties. This potential increase would suggest a significant growth in the number of satellite galaxies hosted within individual dark matter halos during the period when the Universe was still (partially) neutral.

Interestingly, this trend differs from the monotonic predictions of \citet{Bhowmick_2018}, where the \textsc{BlueTides} simulation predicted consistently low satellite fractions during the early stages of reionization. The elevated measurements at $z \sim 7-8$, if confirmed, may reflect the interplay between structure formation and reionization feedback processes not fully captured in current theoretical models. However, we note that our analysis assumes the continued validity of the HOD framework at these high redshifts, where the significantly higher cosmic density and ongoing reionization may introduce systematic effects not accounted for in standard HOD parameterizations.

Should this evolution be confirmed by future studies, it would be well compatible with a boosted efficiency of early star formation in low-mass halos These galaxies would cluster in dense, multi-galaxy systems during cosmic dawn before environmental processes reshaped the galaxy population.

\section{Summary and Perspectives}\label{sec: summary}

In this study we analyzed the clustering of $N_{\rm{g}} \simeq 6500$ LBGs at redshifts $5 \leq z<11$ from the GOODS-South survey conducted by the JADES Collaboration \citep{Bunker_2023,Eisenstein_2023a,Eisenstein_2023b,Hainline_2023,Rieke_2023, D'Eugenio_2024} by adopting a HOD formalism. Leveraging JWST's unprecedented sensitivity, this analysis extends current knowledge on the interplay between galaxies and dark matter halos beyond bright sources to the faint end of the luminosity function and to higher redshifts than previously accessible.

Our major results and conclusions are summarized as follows:

$\bullet$ We found that the redshift evolution of the free HOD parameters is consistent with previous studies with analogous samples. Specifically, both $M_{\rm{min}}[M_{\odot}]$ and $M_{\rm{sat}}[M_{\odot}]$ show a decreasing trend with redshift, suggesting that high-redshift galaxies are hosted by significantly less massive dark matter halos, as also indicated by the $M_{\rm{h}}[M_{\odot}]$ measurements (e.g.,\citealt{Hamana_2004, Conroy_2006, Harikane_2016, Harikane_2021,Paquereau_2025,Shuntov_2025}).

$\bullet$ We studied high-redshift LBGs in previously unexplored regions of the luminosity-redshift parameter space, obtaining new results on galaxy clustering measurements and presenting the galaxy bias evolution from redshift $z=5.5$ with $b_{\rm{g}}^{\rm{eff}}=5.64_{-0.13}^{+0.10}$ to $z=10.6$ with a galaxy bias measurement of $b_{\rm{g}}^{\rm{eff}}=8.13_{-0.02}^{+0.04}$, supporting previous results that show an increasing trend in galaxy bias with redshift (e.g.,\citealt{Tormen_1999, Overzier_2006,Barone_2014,Park_2017,Qiu_2018,Harikane_2021,Dalmasso_2023,Dalmasso_2024, Paquereau_2025, Shuntov_2025}). Notably, our measurement at $z=10.6$ represents the highest redshift constraint to date for LBG clustering.

$\bullet$ We measured rapid change with redshift of the characteristic dark matter halo mass $M_{\rm{h}}$, decreasing by more than one order of magnitude from $\log_{10}(M_{\rm{h}}/M_{\odot}) = 11.45_{-0.04}^{+0.02}$ at $z = 5.5$ to $\log_{10}(M_{\rm{h}}/M_{\odot}) = 10.12_{-0.06}^{+0.08}$ at $z = 10.6$. This evolutionary trend shows that galaxy populations with similar average UV magnitudes are hosted by increasingly less massive dark matter halos at higher redshifts, with the stellar light-to-halo mass ratio rising by roughly a factor of ten from $z \sim 6$ to $z \sim 10$.

$\bullet$ We found an increasing evolution of the galaxy satellite fraction $f_{\rm{sat}}$ with respect to redshift. While satellite galaxies comprise less than $1\%$ of the population at $z\sim5-6$, our measurements suggest this fraction may increase during reionization, reaching values as high as $\sim28\%$ at $z\sim8.5$, though with substantial uncertainties at the highest redshifts. If confirmed, this would suggest highly efficient galaxy formation in dense environments during cosmic dawn, with early galaxies preferentially forming in multi-galaxy systems before environmental processes reduced satellite fractions at later times.\\

By probing the relationship between galaxies and dark matter across cosmic dawn, this work provides new insights into the physical processes that may have governed the earliest phases of galaxy assembly, the potential role of environmental effects in shaping primordial galaxy populations, and the transition from the initial conditions of structure formation to the mature cosmic web we observe today.

\section*{Acknowledgements}
We thank the anonymous referee for useful suggestions and
comments that have improved the manuscript. This research was supported by the Australian Research Council Center of Excellence for All Sky Astrophysics in 3 Dimensions (ASTRO 3D), through project number CE170100013. This research was supported in part by The Dr Albert Shimmins Fund through the Albert Shimmins Postgraduate Writing Up Award (University of Melbourne).

\section*{Data Availability }
The data used to conduct the analysis are from "The JWST Advanced Deep Extragalactic Survey (JADES)" \citep{Bunker_2023,Eisenstein_2023a,Hainline_2023,Rieke_2023, Eisenstein_2023b,D'Eugenio_2024}.


\bibliographystyle{mnras}
\bibliography{main} 

@ARTICLE{Jose_2017,
       author = {{Jose}, Charles and {Baugh}, Carlton M. and {Lacey}, Cedric G. and {Subramanian}, Kandaswamy},
        title = "{Understanding the non-linear clustering of high-redshift galaxies}",
      journal = {\mnras},
         year = 2017,
        month = aug,
       volume = {469},
       number = {4},
        pages = {4428-4436},
          doi = {10.1093/mnras/stx1014},
archivePrefix = {arXiv},
       eprint = {1702.00853},
 primaryClass = {astro-ph.CO},
       adsurl = {https://ui.adsabs.harvard.edu/abs/2017MNRAS.469.4428J}
}

@ARTICLE{Jose_2016,
       author = {{Jose}, Charles and {Lacey}, Cedric G. and {Baugh}, Carlton M.},
        title = "{The clustering of dark matter haloes: scale-dependent bias on quasi-linear scales}",
      journal = {\mnras},
         year = 2016,
        month = nov,
       volume = {463},
       number = {1},
        pages = {270-281},
          doi = {10.1093/mnras/stw1702},
archivePrefix = {arXiv},
       eprint = {1509.06715},
 primaryClass = {astro-ph.CO},
       adsurl = {https://ui.adsabs.harvard.edu/abs/2016MNRAS.463..270J}
}

@ARTICLE{Oke_1983,
       author = {{Oke}, J.~B. and {Gunn}, J.~E.},
        title = "{Secondary standard stars for absolute spectrophotometry.}",
      journal = {\apj},
         year = 1983,
        month = mar,
       volume = {266},
        pages = {713-717},
          doi = {10.1086/160817},
       adsurl = {https://ui.adsabs.harvard.edu/abs/1983ApJ...266..713O}
}

@ARTICLE{Landy_1993,
       author = {{Landy}, Stephen D. and {Szalay}, Alexander S.},
        title = "{Bias and Variance of Angular Correlation Functions}",
      journal = {\apj},
         year = 1993,
        month = jul,
       volume = {412},
        pages = {64},
          doi = {10.1086/172900},
       adsurl = {https://ui.adsabs.harvard.edu/abs/1993ApJ...412...64L}
}

@article{Peacock_1991,
    author = {Peacock, J. A. and Nicholson, D.},
    title = "{The large-scale clustering of radio galaxies}",
    journal = {Monthly Notices of the Royal Astronomical Society},
    volume = {253},
    number = {2},
    pages = {307-319},
    year = {1991},
    month = {11},
    issn = {0035-8711},
    doi = {10.1093/mnras/253.2.307},
    url = {https://doi.org/10.1093/mnras/253.2.307},
    eprint = {https://academic.oup.com/mnras/article-pdf/253/2/307/32290448/mnras253-0307.pdf},
}

@article{Overzier_2006,
 author = {Overzier, Roderik A. and Bouwens, Rychard J. and Illingworth, Garth D. and Franx, Marijn},
 doi = {10.1086/507678},
 journal = {The Astrophysical Journal},
 note = {https://iopscience.iop.org/article/10.1086/507678/pdf},
 number = {1},
 pages = {l5-l8},
 title = {Clustering of i 775 Dropout Galaxies at z ~ 6 in GOODS and the UDF},
 url = {https://app.dimensions.ai/details/publication/pub.1058785924},
 volume = {648},
 year = {2006}
}

@article{Barone_2014,
 author = {Barone-Nugent, R. L. and Trenti, M. and Wyithe, J. S. B. and Bouwens, R. J. and Oesch, P. A. and Illingworth, G. D. and Carollo, C. M. and Su, J. and Stiavelli, M. and Labbe, I. and van Dokkum, P. G.},
 doi = {10.1088/0004-637x/793/1/17},
 journal = {The Astrophysical Journal},
 note = {http://arxiv.org/pdf/1407.7316},
 number = {1},
 pages = {17},
 title = {MEASUREMENT OF GALAXY CLUSTERING AT z ∼ 7.2 AND THE EVOLUTION OF GALAXY BIAS FROM 3.8 &lt; z &lt; 8 IN THE XDF, GOODS-S, AND GOODS-N},
 url = {https://app.dimensions.ai/details/publication/pub.1015996094},
 volume = {793},
 year = {2014}
}

@ARTICLE{Dalmasso_2023,
       author = {{Dalmasso}, Nicol{\`o} and {Trenti}, Michele and {Leethochawalit}, Nicha},
        title = "{Galaxy clustering measurements out to redshift z {\ensuremath{\sim}} 8 from Hubble Legacy Fields}",
      journal = {\mnras},
         year = 2024,
        month = feb,
       volume = {528},
       number = {1},
        pages = {898-908},
          doi = {10.1093/mnras/stad3901},
archivePrefix = {arXiv},
       eprint = {2312.12329},
 primaryClass = {astro-ph.GA},
       adsurl = {https://ui.adsabs.harvard.edu/abs/2024MNRAS.528..898D}
}

@article{Adelberger_2005,
 author = {Adelberger, Kurt L. and Steidel, Charles C. and Pettini, Max and Shapley, Alice E. and Reddy, Naveen A. and Erb, Dawn K.},
 doi = {10.1086/426580},
 journal = {The Astrophysical Journal},
 number = {2},
 pages = {697-713},
 title = {The Spatial Clustering of Star‐forming Galaxies at Redshifts 1.4 ≲ z ≲ 3.5},
 url = {https://app.dimensions.ai/details/publication/pub.1058714752},
 volume = {619},
 year = {2005}
}

@BOOK{Peebles_1980,
       author = {{Peebles}, P.~J.~E.},
        title = "{The large-scale structure of the universe}",
         year = 1980,
       adsurl = {https://ui.adsabs.harvard.edu/abs/1980lssu.book.....P}
}

@article{Leethochawalit_2022,
  title={A quantitative assessment of completeness correction methods and public release of a versatile simulation code},
  author={Leethochawalit, Nicha and Trenti, Michele and Morishita, Takahiro and Roberts-Borsani, Guido and Treu, Tommaso},
  journal={Monthly Notices of the Royal Astronomical Society},
  volume={509},
  number={4},
  pages={5836--5857},
  year={2022},
  publisher={Oxford University Press}
}

@ARTICLE{Williams2018,
       author = {{Williams}, Christina C. and {Curtis-Lake}, Emma and {Hainline}, Kevin N. and {Chevallard}, Jacopo and {Robertson}, Brant E. and {Charlot}, Stephane and {Endsley}, Ryan and {Stark}, Daniel P. and {Willmer}, Christopher N.~A. and {Alberts}, Stacey and {Amorin}, Ricardo and {Arribas}, Santiago and {Baum}, Stefi and {Bunker}, Andrew and {Carniani}, Stefano and {Crandall}, Sara and {Egami}, Eiichi and {Eisenstein}, Daniel J. and {Ferruit}, Pierre and {Husemann}, Bernd and {Maseda}, Michael V. and {Maiolino}, Roberto and {Rawle}, Timothy D. and {Rieke}, Marcia and {Smit}, Renske and {Tacchella}, Sandro and {Willott}, Chris J.},
        title = "{The JWST Extragalactic Mock Catalog: Modeling Galaxy Populations from the UV through the Near-IR over 13 Billion Years of Cosmic History}",
      journal = {\apjs},
         year = 2018,
        month = jun,
       volume = {236},
       number = {2},
          eid = {33},
        pages = {33},
          doi = {10.3847/1538-4365/aabcbb},
archivePrefix = {arXiv},
       eprint = {1802.05272},
 primaryClass = {astro-ph.GA},
       adsurl = {https://ui.adsabs.harvard.edu/abs/2018ApJS..236...33W}
}

@ARTICLE{Rieke_2023,
       author = {{Rieke}, Marcia J. and {Robertson}, Brant E. and {Tacchella}, Sandro and {Hainline}, Kevin and {Johnson}, Benjamin D. and {Hausan}, Ryan and {Ji}, Zhiyuan and {Willmer}, Christopher N.~A. and {Eisenstein}, Daniel J. and {Pusk{\`a}s}, D{\`a}vid and {Alberts}, Stacey and {Arribas}, Santiago and {Baker}, William M. and {Baum}, Stefi and {Bhatawdekar}, Rachana and {Bonaventura}, Nina and {Boyett}, Kit and {Bunker}, Andrew and {Cameron}, Alex J. and {Carniani}, Stefano and {Charlot}, Stephane and {Chevallard}, Jacopo and {Chen}, Zuyi and {Curti}, Mirko and {Curtis-Lake}, Emma and {Danhaive}, A. Lola and {DeCoursey}, Christa and {Dressler}, Alan and {Egami}, Eiichi and {Endsley}, Ryan and {Helton}, Jakob M. and {Hviding}, Raphael E. and {Kumari}, Nimisha and {Looser}, Tobias and {Lyu}, Jianwei and {Maiolino}, Roberto and {Maseda}, Michael V. and {Nelson}, Erica J. and {Rieke}, George and {Rix}, Hans-Walter and {Sandles}, Lester and {Saxena}, Aayush and {Sharpe}, Katherine and {Shivaei}, Irene and {Skarbinski}, Maya and {Smit}, Renske and {Stark}, Daniel P. and {Stone}, Meredith and {Suess}, Katherine A. and {Sun}, Fengwu and {Topping}, Michael and {Uebler}, Hannah and {Villanueva}, Natalia C. and {Wallace}, Imaan B. and {Williams}, Christina C. and {Willott}, Chris and {Whitler}, Lily and {Witstok}, Joris and {Woodrum}, Charity},
        title = "{JADES Initial Data Release for the Hubble Ultra Deep Field: Revealing the Faint Infrared Sky with Deep JWST NIRCam Imaging}",
      journal = {arXiv e-prints},
         year = 2023,
        month = jun,
          eid = {arXiv:2306.02466},
        pages = {arXiv:2306.02466},
          doi = {10.48550/arXiv.2306.02466},
archivePrefix = {arXiv},
       eprint = {2306.02466},
 primaryClass = {astro-ph.GA},
       adsurl = {https://ui.adsabs.harvard.edu/abs/2023arXiv230602466R}
}

@article{Bouwens_2021,
doi = {10.3847/1538-3881/abf83e},
url = {https://dx.doi.org/10.3847/1538-3881/abf83e},
year = {2021},
month = {jul},
publisher = {The American Astronomical Society},
volume = {162},
number = {2},
pages = {47},
author = {R. J. Bouwens and P. A. Oesch and M. Stefanon and G. Illingworth and I. Labbé and N. Reddy and H. Atek and M. Montes and R. Naidu and T. Nanayakkara and E. Nelson and S. Wilkins},
title = {New Determinations of the UV Luminosity Functions from z ∼ 9 to 2 Show a Remarkable Consistency with Halo Growth and a Constant Star Formation Efficiency},
journal = {The Astronomical Journal}
}

@ARTICLE{Ade_Planck_2015,
       author = {{Planck Collaboration} and {Ade}, P.~A.~R. and {Aghanim}, N. and {Arnaud}, M. and {Ashdown}, M. and {Aumont}, J. and {Baccigalupi}, C. and {Banday}, A.~J. and {Barreiro}, R.~B. and {Bartlett}, J.~G. and {Bartolo}, N. and {Battaner}, E. and {Battye}, R. and {Benabed}, K. and {Beno{\^\i}t}, A. and {Benoit-L{\'e}vy}, A. and {Bernard}, J. -P. and {Bersanelli}, M. and {Bielewicz}, P. and {Bock}, J.~J. and {Bonaldi}, A. and {Bonavera}, L. and {Bond}, J.~R. and {Borrill}, J. and {Bouchet}, F.~R. and {Boulanger}, F. and {Bucher}, M. and {Burigana}, C. and {Butler}, R.~C. and {Calabrese}, E. and {Cardoso}, J. -F. and {Catalano}, A. and {Challinor}, A. and {Chamballu}, A. and {Chary}, R. -R. and {Chiang}, H.~C. and {Chluba}, J. and {Christensen}, P.~R. and {Church}, S. and {Clements}, D.~L. and {Colombi}, S. and {Colombo}, L.~P.~L. and {Combet}, C. and {Coulais}, A. and {Crill}, B.~P. and {Curto}, A. and {Cuttaia}, F. and {Danese}, L. and {Davies}, R.~D. and {Davis}, R.~J. and {de Bernardis}, P. and {de Rosa}, A. and {de Zotti}, G. and {Delabrouille}, J. and {D{\'e}sert}, F. -X. and {Di Valentino}, E. and {Dickinson}, C. and {Diego}, J.~M. and {Dolag}, K. and {Dole}, H. and {Donzelli}, S. and {Dor{\'e}}, O. and {Douspis}, M. and {Ducout}, A. and {Dunkley}, J. and {Dupac}, X. and {Efstathiou}, G. and {Elsner}, F. and {En{\ss}lin}, T.~A. and {Eriksen}, H.~K. and {Farhang}, M. and {Fergusson}, J. and {Finelli}, F. and {Forni}, O. and {Frailis}, M. and {Fraisse}, A.~A. and {Franceschi}, E. and {Frejsel}, A. and {Galeotta}, S. and {Galli}, S. and {Ganga}, K. and {Gauthier}, C. and {Gerbino}, M. and {Ghosh}, T. and {Giard}, M. and {Giraud-H{\'e}raud}, Y. and {Giusarma}, E. and {Gjerl{\o}w}, E. and {Gonz{\'a}lez-Nuevo}, J. and {G{\'o}rski}, K.~M. and {Gratton}, S. and {Gregorio}, A. and {Gruppuso}, A. and {Gudmundsson}, J.~E. and {Hamann}, J. and {Hansen}, F.~K. and {Hanson}, D. and {Harrison}, D.~L. and {Helou}, G. and {Henrot-Versill{\'e}}, S. and {Hern{\'a}ndez-Monteagudo}, C. and {Herranz}, D. and {Hildebrandt}, S.~R. and {Hivon}, E. and {Hobson}, M. and {Holmes}, W.~A. and {Hornstrup}, A. and {Hovest}, W. and {Huang}, Z. and {Huffenberger}, K.~M. and {Hurier}, G. and {Jaffe}, A.~H. and {Jaffe}, T.~R. and {Jones}, W.~C. and {Juvela}, M. and {Keih{\"a}nen}, E. and {Keskitalo}, R. and {Kisner}, T.~S. and {Kneissl}, R. and {Knoche}, J. and {Knox}, L. and {Kunz}, M. and {Kurki-Suonio}, H. and {Lagache}, G. and {L{\"a}hteenm{\"a}ki}, A. and {Lamarre}, J. -M. and {Lasenby}, A. and {Lattanzi}, M. and {Lawrence}, C.~R. and {Leahy}, J.~P. and {Leonardi}, R. and {Lesgourgues}, J. and {Levrier}, F. and {Lewis}, A. and {Liguori}, M. and {Lilje}, P.~B. and {Linden-V{\o}rnle}, M. and {L{\'o}pez-Caniego}, M. and {Lubin}, P.~M. and {Mac{\'\i}as-P{\'e}rez}, J.~F. and {Maggio}, G. and {Maino}, D. and {Mandolesi}, N. and {Mangilli}, A. and {Marchini}, A. and {Maris}, M. and {Martin}, P.~G. and {Martinelli}, M. and {Mart{\'\i}nez-Gonz{\'a}lez}, E. and {Masi}, S. and {Matarrese}, S. and {McGehee}, P. and {Meinhold}, P.~R. and {Melchiorri}, A. and {Melin}, J. -B. and {Mendes}, L. and {Mennella}, A. and {Migliaccio}, M. and {Millea}, M. and {Mitra}, S. and {Miville-Desch{\^e}nes}, M. -A. and {Moneti}, A. and {Montier}, L. and {Morgante}, G. and {Mortlock}, D. and {Moss}, A. and {Munshi}, D. and {Murphy}, J.~A. and {Naselsky}, P. and {Nati}, F. and {Natoli}, P. and {Netterfield}, C.~B. and {N{\o}rgaard-Nielsen}, H.~U. and {Noviello}, F. and {Novikov}, D. and {Novikov}, I. and {Oxborrow}, C.~A. and {Paci}, F. and {Pagano}, L. and {Pajot}, F. and {Paladini}, R. and {Paoletti}, D. and {Partridge}, B. and {Pasian}, F. and {Patanchon}, G. and {Pearson}, T.~J. and {Perdereau}, O. and {Perotto}, L. and {Perrotta}, F. and {Pettorino}, V. and {Piacentini}, F. and {Piat}, M. and {Pierpaoli}, E. and {Pietrobon}, D. and {Plaszczynski}, S. and {Pointecouteau}, E. and {Polenta}, G. and {Popa}, L. and {Pratt}, G.~W. and {Pr{\'e}zeau}, G. and {Prunet}, S. and {Puget}, J. -L. and {Rachen}, J.~P. and {Reach}, W.~T. and {Rebolo}, R. and {Reinecke}, M. and {Remazeilles}, M. and {Renault}, C. and {Renzi}, A. and {Ristorcelli}, I. and {Rocha}, G. and {Rosset}, C. and {Rossetti}, M. and {Roudier}, G. and {Rouill{\'e} d'Orfeuil}, B. and {Rowan-Robinson}, M. and {Rubi{\~n}o-Mart{\'\i}n}, J.~A. and {Rusholme}, B. and {Said}, N. and {Salvatelli}, V. and {Salvati}, L. and {Sandri}, M. and {Santos}, D. and {Savelainen}, M. and {Savini}, G. and {Scott}, D. and {Seiffert}, M.~D. and {Serra}, P. and {Shellard}, E.~P.~S. and {Spencer}, L.~D. and {Spinelli}, M. and {Stolyarov}, V. and {Stompor}, R. and {Sudiwala}, R. and {Sunyaev}, R. and {Sutton}, D. and {Suur-Uski}, A. -S. and {Sygnet}, J. -F. and {Tauber}, J.~A. and {Terenzi}, L. and {Toffolatti}, L. and {Tomasi}, M. and {Tristram}, M. and {Trombetti}, T. and {Tucci}, M. and {Tuovinen}, J. and {T{\"u}rler}, M. and {Umana}, G. and {Valenziano}, L. and {Valiviita}, J. and {Van Tent}, F. and {Vielva}, P. and {Villa}, F. and {Wade}, L.~A. and {Wandelt}, B.~D. and {Wehus}, I.~K. and {White}, M. and {White}, S.~D.~M. and {Wilkinson}, A. and {Yvon}, D. and {Zacchei}, A. and {Zonca}, A.},
        title = "{Planck 2015 results. XIII. Cosmological parameters}",
      journal = {\aap},
         year = 2016,
        month = sep,
       volume = {594},
          eid = {A13},
        pages = {A13},
          doi = {10.1051/0004-6361/201525830},
archivePrefix = {arXiv},
       eprint = {1502.01589},
 primaryClass = {astro-ph.CO},
       adsurl = {https://ui.adsabs.harvard.edu/abs/2016A&A...594A..13P}
}

@ARTICLE{Bunker_2023,
       author = {{Bunker}, Andrew J. and {Cameron}, Alex J. and {Curtis-Lake}, Emma and {Jakobsen}, Peter and {Carniani}, Stefano and {Curti}, Mirko and {Witstok}, Joris and {Maiolino}, Roberto and {D'Eugenio}, Francesco and {Looser}, Tobias J. and {Willott}, Chris and {Bonaventura}, Nina and {Hainline}, Kevin and {Uebler}, Hannah and {Willmer}, Christopher N.~A. and {Saxena}, Aayush and {Smit}, Renske and {Alberts}, Stacey and {Arribas}, Santiago and {Baker}, William M. and {Baum}, Stefi and {Bhatawdekar}, Rachana and {Bowler}, Rebecca A.~A. and {Boyett}, Kristan and {Charlot}, Stephane and {Chen}, Zuyi and {Chevallard}, Jacopo and {Circosta}, Chiara and {DeCoursey}, Christa and {de Graaff}, Anna and {Egami}, Eiichi and {Eisenstein}, Daniel J. and {Endsley}, Ryan and {Ferruit}, Pierre and {Giardino}, Giovanna and {Hausen}, Ryan and {Helton}, Jakob M. and {Hviding}, Raphael E. and {Ji}, Zhiyuan and {Johnson}, Benjamin D. and {Jones}, Gareth C. and {Kumari}, Nimisha and {Laseter}, Isaac and {Luetzgendorf}, Nora and {Maseda}, Michael V. and {Nelson}, Erica and {Parlanti}, Eleonora and {Perna}, Michele and {Rawle}, Tim and {Rix}, Hans-Walter and {Rieke}, Marcia and {Robertson}, Brant and {Rodriguez Del Pino}, Bruno and {Sandles}, Lester and {Scholtz}, Jan and {Sharpe}, Katherine and {Skarbinski}, Maya and {Stark}, Daniel P. and {Sun}, Fengwu and {Tacchella}, Sandro and {Topping}, Michael W. and {Villanueva}, Natalia C. and {Wallace}, Imaan E.~B. and {Williams}, Christina C. and {Woodrum}, Charity},
        title = "{JADES NIRSpec Initial Data Release for the Hubble Ultra Deep Field: Redshifts and Line Fluxes of Distant Galaxies from the Deepest JWST Cycle 1 NIRSpec Multi-Object Spectroscopy}",
      journal = {arXiv e-prints},
         year = 2023,
        month = jun,
          eid = {arXiv:2306.02467},
        pages = {arXiv:2306.02467},
          doi = {10.48550/arXiv.2306.02467},
archivePrefix = {arXiv},
       eprint = {2306.02467},
 primaryClass = {astro-ph.GA},
       adsurl = {https://ui.adsabs.harvard.edu/abs/2023arXiv230602467B}
}

@ARTICLE{Eisenstein_2023a,
       author = {{Eisenstein}, Daniel J. and {Willott}, Chris and {Alberts}, Stacey and {Arribas}, Santiago and {Bonaventura}, Nina and {Bunker}, Andrew J. and {Cameron}, Alex J. and {Carniani}, Stefano and {Charlot}, Stephane and {Curtis-Lake}, Emma and {D'Eugenio}, Francesco and {Endsley}, Ryan and {Ferruit}, Pierre and {Giardino}, Giovanna and {Hainline}, Kevin and {Hausen}, Ryan and {Jakobsen}, Peter and {Johnson}, Benjamin D. and {Maiolino}, Roberto and {Rieke}, Marcia and {Rieke}, George and {Rix}, Hans-Walter and {Robertson}, Brant and {Stark}, Daniel P. and {Tacchella}, Sandro and {Williams}, Christina C. and {Willmer}, Christopher N.~A. and {Baker}, William M. and {Baum}, Stefi and {Bhatawdekar}, Rachana and {Boyett}, Kristan and {Chen}, Zuyi and {Chevallard}, Jacopo and {Circosta}, Chiara and {Curti}, Mirko and {Danhaive}, A. Lola and {DeCoursey}, Christa and {de Graaff}, Anna and {Dressler}, Alan and {Egami}, Eiichi and {Helton}, Jakob M. and {Hviding}, Raphael E. and {Ji}, Zhiyuan and {Jones}, Gareth C. and {Kumari}, Nimisha and {L{\"u}tzgendorf}, Nora and {Laseter}, Isaac and {Looser}, Tobias J. and {Lyu}, Jianwei and {Maseda}, Michael V. and {Nelson}, Erica and {Parlanti}, Eleonora and {Perna}, Michele and {Pusk{\'a}s}, D{\'a}vid and {Rawle}, Tim and {Rodr{\'\i}guez Del Pino}, Bruno and {Sandles}, Lester and {Saxena}, Aayush and {Scholtz}, Jan and {Sharpe}, Katherine and {Shivaei}, Irene and {Silcock}, Maddie S. and {Simmonds}, Charlotte and {Skarbinski}, Maya and {Smit}, Renske and {Stone}, Meredith and {Suess}, Katherine A. and {Sun}, Fengwu and {Tang}, Mengtao and {Topping}, Michael W. and {{\"U}bler}, Hannah and {Villanueva}, Natalia C. and {Wallace}, Imaan E.~B. and {Whitler}, Lily and {Witstok}, Joris and {Woodrum}, Charity},
        title = "{Overview of the JWST Advanced Deep Extragalactic Survey (JADES)}",
      journal = {arXiv e-prints},
         year = 2023,
        month = jun,
          eid = {arXiv:2306.02465},
        pages = {arXiv:2306.02465},
          doi = {10.48550/arXiv.2306.02465},
archivePrefix = {arXiv},
       eprint = {2306.02465},
 primaryClass = {astro-ph.GA},
       adsurl = {https://ui.adsabs.harvard.edu/abs/2023arXiv230602465E}
}

@ARTICLE{Hainline_2023,
       author = {{Hainline}, Kevin N. and {Johnson}, Benjamin D. and {Robertson}, Brant and {Tacchella}, Sandro and {Helton}, Jakob M. and {Sun}, Fengwu and {Eisenstein}, Daniel J. and {Simmonds}, Charlotte and {Topping}, Michael W. and {Whitler}, Lily and {Willmer}, Christopher N.~A. and {Rieke}, Marcia and {Suess}, Katherine A. and {Hviding}, Raphael E. and {Cameron}, Alex J. and {Alberts}, Stacey and {Baker}, William M. and {Bhatawdekar}, Rachana and {Boyett}, Kristan and {Bunker}, Andrew J. and {Carniani}, Stefano and {Charlot}, Stephane and {Chen}, Zuyi and {Curti}, Mirko and {Curtis-Lake}, Emma and {D'Eugenio}, Francesco and {Egami}, Eiichi and {Endsley}, Ryan and {Hausen}, Ryan and {Ji}, Zhiyuan and {Looser}, Tobias J. and {Lyu}, Jianwei and {Maiolino}, Roberto and {Nelson}, Erica and {Puskas}, David and {Rawle}, Tim and {Sandles}, Lester and {Saxena}, Aayush and {Smit}, Renske and {Stark}, Daniel P. and {Williams}, Christina C. and {Willott}, Chris and {Witstok}, Joris},
        title = "{The Cosmos in its Infancy: JADES Galaxy Candidates at z > 8 in GOODS-S and GOODS-N}",
      journal = {arXiv e-prints},
         year = 2023,
        month = jun,
          eid = {arXiv:2306.02468},
        pages = {arXiv:2306.02468},
          doi = {10.48550/arXiv.2306.02468},
archivePrefix = {arXiv},
       eprint = {2306.02468},
 primaryClass = {astro-ph.GA},
       adsurl = {https://ui.adsabs.harvard.edu/abs/2023arXiv230602468H}
}

@ARTICLE{Eisenstein_2023b,
       author = {{Eisenstein}, Daniel J. and {Johnson}, Benjamin D. and {Robertson}, Brant and {Tacchella}, Sandro and {Hainline}, Kevin and {Jakobsen}, Peter and {Maiolino}, Roberto and {Bonaventura}, Nina and {Bunker}, Andrew J. and {Cameron}, Alex J. and {Cargile}, Phillip A. and {Curtis-Lake}, Emma and {Hausen}, Ryan and {Pusk{\'a}s}, D{\'a}vid and {Rieke}, Marcia and {Sun}, Fengwu and {Willmer}, Christopher N.~A. and {Willott}, Chris and {Alberts}, Stacey and {Arribas}, Santiago and {Baker}, William M. and {Baum}, Stefi and {Bhatawdekar}, Rachana and {Carniani}, Stefano and {Charlot}, Stephane and {Chen}, Zuyi and {Chevallard}, Jacopo and {Curti}, Mirko and {DeCoursey}, Christa and {D'Eugenio}, Francesco and {de Graaff}, Anna and {Egami}, Eiichi and {Helton}, Jakob M. and {Ji}, Zhiyuan and {Jones}, Gareth C. and {Kumari}, Nimisha and {L{\"u}tzgendorf}, Nora and {Laseter}, Isaac and {Looser}, Tobias J. and {Lyu}, Jianwei and {Maseda}, Michael V. and {Nelson}, Erica and {Parlanti}, Eleonora and {Rauscher}, Bernard J. and {Rawle}, Tim and {Rieke}, George and {Rix}, Hans-Walter and {Rujopakarn}, Wiphu and {Sandles}, Lester and {Saxena}, Aayush and {Scholtz}, Jan and {Sharpe}, Katherine and {Shivaei}, Irene and {Simmonds}, Charlotte and {Smit}, Renske and {Topping}, Michael W. and {{\"U}bler}, Hannah and {Venturi}, Giacomo and {Williams}, Christina C. and {Witstok}, Joris and {Woodrum}, Charity},
        title = "{The JADES Origins Field: A New JWST Deep Field in the JADES Second NIRCam Data Release}",
      journal = {arXiv e-prints},
         year = 2023,
        month = oct,
          eid = {arXiv:2310.12340},
        pages = {arXiv:2310.12340},
          doi = {10.48550/arXiv.2310.12340},
archivePrefix = {arXiv},
       eprint = {2310.12340},
 primaryClass = {astro-ph.GA},
       adsurl = {https://ui.adsabs.harvard.edu/abs/2023arXiv231012340E}
}

@ARTICLE{Groth_1977,
       author = {{Groth}, E.~J. and {Peebles}, P.~J.~E.},
        title = "{Statistical analysis of catalogs of extragalactic objects. VII. Two- and three-point correlation functions for the high-resolution Shane-Wirtanen catalog of galaxies.}",
      journal = {\apj},
         year = 1977,
        month = oct,
       volume = {217},
        pages = {385-405},
          doi = {10.1086/155588},
       adsurl = {https://ui.adsabs.harvard.edu/abs/1977ApJ...217..385G}
}

@ARTICLE{Kravtsov_2004,
       author = {{Kravtsov}, Andrey V. and {Berlind}, Andreas A. and {Wechsler}, Risa H. and {Klypin}, Anatoly A. and {Gottl{\"o}ber}, Stefan and {Allgood}, Brandon and {Primack}, Joel R.},
        title = "{The Dark Side of the Halo Occupation Distribution}",
      journal = {\apj},
         year = 2004,
        month = jul,
       volume = {609},
       number = {1},
        pages = {35-49},
          doi = {10.1086/420959},
archivePrefix = {arXiv},
       eprint = {astro-ph/0308519},
 primaryClass = {astro-ph},
       adsurl = {https://ui.adsabs.harvard.edu/abs/2004ApJ...609...35K}
}

@ARTICLE{Conroy_2006,
       author = {{Conroy}, Charlie and {Wechsler}, Risa H. and {Kravtsov}, Andrey V.},
        title = "{Modeling Luminosity-dependent Galaxy Clustering through Cosmic Time}",
      journal = {\apj},
         year = 2006,
        month = aug,
       volume = {647},
       number = {1},
        pages = {201-214},
          doi = {10.1086/503602},
archivePrefix = {arXiv},
       eprint = {astro-ph/0512234},
 primaryClass = {astro-ph},
       adsurl = {https://ui.adsabs.harvard.edu/abs/2006ApJ...647..201C}
}

@article{Ishikawa_2017,
  title={The Subaru HSC Galaxy Clustering with Photometric Redshift. I. Dark Halo Masses versus Baryonic Properties of Galaxies at 0.3 ≤ z ≤ 1.4},
  author={Shogo Ishikawa and Nobunari Kashikawa and Masayuki Tanaka and Jean Coupon and Alexie Leauthaud and Jun Toshikawa and Kohei Ichikawa and Taira Oogi and Hisakazu Uchiyama and Yuu Niino and Atsushi J. Nishizawa},
  journal={The Astrophysical Journal},
  year={2019},
  volume={904},
  url={https://api.semanticscholar.org/CorpusID:209324191}
}

@article{Harikane_2021,
  title={GOLDRUSH. IV. Luminosity Functions and Clustering Revealed with~ 4,000,000 Galaxies at z~ 2--7: Galaxy--AGN Transition, Star Formation Efficiency, and Implication for Evolution at z> 10},
  author={Harikane, Yuichi and Ono, Yoshiaki and Ouchi, Masami and Liu, Chengze and Sawicki, Marcin and Shibuya, Takatoshi and Behroozi, Peter S and He, Wanqiu and Shimasaku, Kazuhiro and Arnouts, Stephane and others},
  journal={The Astrophysical Journal Supplement Series},
  volume={259},
  number={1},
  pages={20},
  year={2022},
  publisher={IOP Publishing}
}

@article{Harikane_2016,
doi = {10.3847/0004-637X/821/2/123},
url = {https://dx.doi.org/10.3847/0004-637X/821/2/123},
year = {2016},
month = {apr},
publisher = {The American Astronomical Society},
volume = {821},
number = {2},
pages = {123},
author = {Yuichi Harikane and Masami Ouchi and Yoshiaki Ono and Surhud More and Shun Saito and Yen-Ting Lin and Jean Coupon and Kazuhiro Shimasaku and Takatoshi Shibuya and Paul A. Price and Lihwai Lin and Bau-Ching Hsieh and Masafumi Ishigaki and Yutaka Komiyama and John Silverman and Tadafumi Takata and Hiroko Tamazawa and Jun Toshikawa},
title = {EVOLUTION OF STELLAR-TO-HALO MASS RATIO AT z = 0–7 IDENTIFIED BY CLUSTERING ANALYSIS WITH THE HUBBLE LEGACY IMAGING AND EARLY SUBARU/HYPER SUPRIME-CAM SURVEY DATA},
journal = {The Astrophysical Journal}
}

@article{Hamana_2004,
    author = {Hamana, Takashi and Ouchi, Masami and Shimasaku, Kazuhiro and Kayo, Issha and Suto, Yasushi},
    title = "{Properties of host haloes of Lyman-break galaxies and Lyman α emitters from their number densities and angular clustering}",
    journal = {Monthly Notices of the Royal Astronomical Society},
    volume = {347},
    number = {3},
    pages = {813-823},
    year = {2004},
    month = {01},
    issn = {0035-8711},
    doi = {10.1111/j.1365-2966.2004.07253.x},
    url = {https://doi.org/10.1111/j.1365-2966.2004.07253.x},
    eprint = {https://academic.oup.com/mnras/article-pdf/347/3/813/3350626/347-3-813.pdf},
}

@article{Zehavi_2011,
doi = {10.1088/0004-637X/736/1/59},
url = {https://dx.doi.org/10.1088/0004-637X/736/1/59},
year = {2011},
month = {jul},
publisher = {The American Astronomical Society},
volume = {736},
number = {1},
pages = {59},
author = {Idit Zehavi and Zheng Zheng and David H. Weinberg and Michael R. Blanton and Neta A. Bahcall and Andreas A. Berlind and Jon Brinkmann and Joshua A. Frieman and James E. Gunn and Robert H. Lupton and Robert C. Nichol and Will J. Percival and Donald P. Schneider and Ramin A. Skibba and Michael A. Strauss and Max Tegmark and Donald G. York},
title = {GALAXY CLUSTERING IN THE COMPLETED SDSS REDSHIFT SURVEY: THE DEPENDENCE ON COLOR AND LUMINOSITY},
journal = {The Astrophysical Journal}
}

@ARTICLE{Tormen_1999,
       author = {{Sheth}, Ravi K. and {Tormen}, Giuseppe},
        title = "{Large-scale bias and the peak background split}",
      journal = {\mnras},
         year = 1999,
        month = sep,
       volume = {308},
       number = {1},
        pages = {119-126},
          doi = {10.1046/j.1365-8711.1999.02692.x},
archivePrefix = {arXiv},
       eprint = {astro-ph/9901122},
 primaryClass = {astro-ph},
       adsurl = {https://ui.adsabs.harvard.edu/abs/1999MNRAS.308..119S}
}

@ARTICLE{Qiu_2018,
       author = {{Qiu}, Yisheng and {Wyithe}, J. Stuart B. and {Oesch}, Pascal A. and {Mutch}, Simon J. and {Qin}, Yuxiang and {Labb{\'e}}, Ivo and {Bouwens}, Rychard J. and {Stefanon}, Mauro and {Illingworth}, Garth D.},
        title = "{Dependence of galaxy clustering on UV luminosity and stellar mass at z {\ensuremath{\sim}} 4-7}",
      journal = {\mnras},
         year = 2018,
        month = dec,
       volume = {481},
       number = {4},
        pages = {4885-4894},
          doi = {10.1093/mnras/sty2633},
archivePrefix = {arXiv},
       eprint = {1809.10161},
 primaryClass = {astro-ph.GA},
       adsurl = {https://ui.adsabs.harvard.edu/abs/2018MNRAS.481.4885Q}
}

@article{Mason_2023,
  title={The brightest galaxies at cosmic dawn},
  author={Mason, Charlotte A and Trenti, Michele and Treu, Tommaso},
  journal={Monthly Notices of the Royal Astronomical Society},
  volume={521},
  number={1},
  pages={497--503},
  year={2023},
  publisher={Oxford University Press}
}

@ARTICLE{Boyett_2024,
       author = {{Boyett}, Kristan and {Trenti}, Michele and {Leethochawalit}, Nicha and {Calabr{\'o}}, Antonello and {Metha}, Benjamin and {Roberts-Borsani}, Guido and {Dalmasso}, Nicol{\'o} and {Yang}, Lilan and {Santini}, Paola and {Treu}, Tommaso and {Jones}, Tucker and {Henry}, Alaina and {Mason}, Charlotte A. and {Morishita}, Takahiro and {Nanayakkara}, Themiya and {Roy}, Namrata and {Wang}, Xin and {Fontana}, Adriano and {Merlin}, Emiliano and {Castellano}, Marco and {Paris}, Diego and {Brada{\v{c}}}, Maru{\v{s}}a and {Malkan}, Matt and {Marchesini}, Danilo and {Mascia}, Sara and {Glazebrook}, Karl and {Pentericci}, Laura and {Vanzella}, Eros and {Vulcani}, Benedetta},
        title = "{A massive interacting galaxy 510 million years after the Big Bang}",
      journal = {Nature Astronomy},
         year = 2024,
        month = may,
       volume = {8},
        pages = {657-672},
          doi = {10.1038/s41550-024-02218-7},
archivePrefix = {arXiv},
       eprint = {2303.00306},
 primaryClass = {astro-ph.GA},
       adsurl = {https://ui.adsabs.harvard.edu/abs/2024NatAs...8..657B}
}

@ARTICLE{Labbe_2023,
       author = {{Labb{\'e}}, Ivo and {van Dokkum}, Pieter and {Nelson}, Erica and {Bezanson}, Rachel and {Suess}, Katherine A. and {Leja}, Joel and {Brammer}, Gabriel and {Whitaker}, Katherine and {Mathews}, Elijah and {Stefanon}, Mauro and {Wang}, Bingjie},
        title = "{A population of red candidate massive galaxies  600 Myr after the Big Bang}",
      journal = {\nat},
         year = 2023,
        month = apr,
       volume = {616},
       number = {7956},
        pages = {266-269},
          doi = {10.1038/s41586-023-05786-2},
archivePrefix = {arXiv},
       eprint = {2207.12446},
 primaryClass = {astro-ph.GA},
       adsurl = {https://ui.adsabs.harvard.edu/abs/2023Natur.616..266L}
}

@ARTICLE{Carniani_2024,
       author = {{Carniani}, Stefano and {Hainline}, Kevin and {D'Eugenio}, Francesco and {Eisenstein}, Daniel J. and {Jakobsen}, Peter and {Witstok}, Joris and {Johnson}, Benjamin D. and {Chevallard}, Jacopo and {Maiolino}, Roberto and {Helton}, Jakob M. and {Willott}, Chris and {Robertson}, Brant and {Alberts}, Stacey and {Arribas}, Santiago and {Baker}, William M. and {Bhatawdekar}, Rachana and {Boyett}, Kristan and {Bunker}, Andrew J. and {Cameron}, Alex J. and {Cargile}, Phillip A. and {Charlot}, St{\'e}phane and {Curti}, Mirko and {Curtis-Lake}, Emma and {Egami}, Eiichi and {Giardino}, Giovanna and {Isaak}, Kate and {Ji}, Zhiyuan and {Jones}, Gareth C. and {Kumari}, Nimisha and {Maseda}, Michael V. and {Parlanti}, Eleonora and {P{\'e}rez-Gonz{\'a}lez}, Pablo G. and {Rawle}, Tim and {Rieke}, George and {Rieke}, Marcia and {Del Pino}, Bruno Rodr{\'\i}guez and {Saxena}, Aayush and {Scholtz}, Jan and {Smit}, Renske and {Sun}, Fengwu and {Tacchella}, Sandro and {{\"U}bler}, Hannah and {Venturi}, Giacomo and {Williams}, Christina C. and {Willmer}, Christopher N.~A.},
        title = "{Spectroscopic confirmation of two luminous galaxies at a redshift of 14}",
      journal = {\nat},
         year = 2024,
        month = sep,
       volume = {633},
       number = {8029},
        pages = {318-322},
          doi = {10.1038/s41586-024-07860-9},
archivePrefix = {arXiv},
       eprint = {2405.18485},
 primaryClass = {astro-ph.GA},
       adsurl = {https://ui.adsabs.harvard.edu/abs/2024Natur.633..318C}
}

@ARTICLE{Arrabal_2023,
       author = {{Arrabal Haro}, Pablo and {Dickinson}, Mark and {Finkelstein}, Steven L. and {Kartaltepe}, Jeyhan S. and {Donnan}, Callum T. and {Burgarella}, Denis and {Carnall}, Adam C. and {Cullen}, Fergus and {Dunlop}, James S. and {Fern{\'a}ndez}, Vital and {Fujimoto}, Seiji and {Jung}, Intae and {Krips}, Melanie and {Larson}, Rebecca L. and {Papovich}, Casey and {P{\'e}rez-Gonz{\'a}lez}, Pablo G. and {Amor{\'\i}n}, Ricardo O. and {Bagley}, Micaela B. and {Buat}, V{\'e}ronique and {Casey}, Caitlin M. and {Chworowsky}, Katherine and {Cohen}, Seth H. and {Ferguson}, Henry C. and {Giavalisco}, Mauro and {Huertas-Company}, Marc and {Hutchison}, Taylor A. and {Kocevski}, Dale D. and {Koekemoer}, Anton M. and {Lucas}, Ray A. and {McLeod}, Derek J. and {McLure}, Ross J. and {Pirzkal}, Norbert and {Seill{\'e}}, Lise-Marie and {Trump}, Jonathan R. and {Weiner}, Benjamin J. and {Wilkins}, Stephen M. and {Zavala}, Jorge A.},
        title = "{Confirmation and refutation of very luminous galaxies in the early Universe}",
      journal = {\nat},
         year = 2023,
        month = oct,
       volume = {622},
       number = {7984},
        pages = {707-711},
          doi = {10.1038/s41586-023-06521-7},
archivePrefix = {arXiv},
       eprint = {2303.15431},
 primaryClass = {astro-ph.GA},
       adsurl = {https://ui.adsabs.harvard.edu/abs/2023Natur.622..707A}
}

@ARTICLE{Ferrara_2023,
       author = {{Ferrara}, Andrea and {Pallottini}, Andrea and {Dayal}, Pratika},
        title = "{On the stunning abundance of super-early, luminous galaxies revealed by JWST}",
      journal = {\mnras},
         year = 2023,
        month = jul,
       volume = {522},
       number = {3},
        pages = {3986-3991},
          doi = {10.1093/mnras/stad1095},
archivePrefix = {arXiv},
       eprint = {2208.00720},
 primaryClass = {astro-ph.GA},
       adsurl = {https://ui.adsabs.harvard.edu/abs/2023MNRAS.522.3986F}
}

@ARTICLE{Naidu_2022,
       author = {{Naidu}, Rohan P. and {Oesch}, Pascal A. and {van Dokkum}, Pieter and {Nelson}, Erica J. and {Suess}, Katherine A. and {Brammer}, Gabriel and {Whitaker}, Katherine E. and {Illingworth}, Garth and {Bouwens}, Rychard and {Tacchella}, Sandro and {Matthee}, Jorryt and {Allen}, Natalie and {Bezanson}, Rachel and {Conroy}, Charlie and {Labbe}, Ivo and {Leja}, Joel and {Leonova}, Ecaterina and {Magee}, Dan and {Price}, Sedona H. and {Setton}, David J. and {Strait}, Victoria and {Stefanon}, Mauro and {Toft}, Sune and {Weaver}, John R. and {Weibel}, Andrea},
        title = "{Two Remarkably Luminous Galaxy Candidates at z {\ensuremath{\approx}} 10-12 Revealed by JWST}",
      journal = {\apjl},
         year = 2022,
        month = nov,
       volume = {940},
       number = {1},
          eid = {L14},
        pages = {L14},
          doi = {10.3847/2041-8213/ac9b22},
archivePrefix = {arXiv},
       eprint = {2207.09434},
 primaryClass = {astro-ph.GA},
       adsurl = {https://ui.adsabs.harvard.edu/abs/2022ApJ...940L..14N}
}

@ARTICLE{Kolchin_2023,
       author = {{Boylan-Kolchin}, Michael},
        title = "{Stress testing {\ensuremath{\Lambda}}CDM with high-redshift galaxy candidates}",
      journal = {Nature Astronomy},
         year = 2023,
        month = jun,
       volume = {7},
        pages = {731-735},
          doi = {10.1038/s41550-023-01937-7},
archivePrefix = {arXiv},
       eprint = {2208.01611},
 primaryClass = {astro-ph.CO},
       adsurl = {https://ui.adsabs.harvard.edu/abs/2023NatAs...7..731B}
}

@ARTICLE{Giavalisco_1998,
       author = {{Giavalisco}, Mauro and {Steidel}, Charles C. and {Adelberger}, Kurt L. and {Dickinson}, Mark E. and {Pettini}, Max and {Kellogg}, Melinda},
        title = "{The Angular Clustering of Lyman-Break Galaxies at Redshift Z approximately 3}",
      journal = {\apj},
         year = 1998,
        month = aug,
       volume = {503},
       number = {2},
        pages = {543-552},
          doi = {10.1086/306027},
archivePrefix = {arXiv},
       eprint = {astro-ph/9802318},
 primaryClass = {astro-ph},
       adsurl = {https://ui.adsabs.harvard.edu/abs/1998ApJ...503..543G}
}

@ARTICLE{Giavalisco_2001,
       author = {{Giavalisco}, Mauro and {Dickinson}, Mark},
        title = "{Clustering Segregation with Ultraviolet Luminosity in Lyman Break Galaxies at z\raisebox{-0.5ex}\textasciitilde3 and Its Implications}",
      journal = {\apj},
         year = 2001,
        month = mar,
       volume = {550},
       number = {1},
        pages = {177-194},
          doi = {10.1086/319715},
archivePrefix = {arXiv},
       eprint = {astro-ph/0012249},
 primaryClass = {astro-ph},
       adsurl = {https://ui.adsabs.harvard.edu/abs/2001ApJ...550..177G}
}

@ARTICLE{Porciani_2002,
       author = {{Porciani}, Cristiano and {Giavalisco}, Mauro},
        title = "{The Clustering Properties of Lyman Break Galaxies at Redshift z\raisebox{-0.5ex}\textasciitilde3}",
      journal = {\apj},
         year = 2002,
        month = jan,
       volume = {565},
       number = {1},
        pages = {24-49},
          doi = {10.1086/324198},
archivePrefix = {arXiv},
       eprint = {astro-ph/0107447},
 primaryClass = {astro-ph},
       adsurl = {https://ui.adsabs.harvard.edu/abs/2002ApJ...565...24P}
}

@ARTICLE{Mo_1996,
       author = {{Mo}, H.~J. and {White}, S.~D.~M.},
        title = "{An analytic model for the spatial clustering of dark matter haloes}",
      journal = {\mnras},
         year = 1996,
        month = sep,
       volume = {282},
       number = {2},
        pages = {347-361},
          doi = {10.1093/mnras/282.2.347},
archivePrefix = {arXiv},
       eprint = {astro-ph/9512127},
 primaryClass = {astro-ph},
       adsurl = {https://ui.adsabs.harvard.edu/abs/1996MNRAS.282..347M}
}

@ARTICLE{Ouchi_2004,
       author = {{Ouchi}, Masami and {Shimasaku}, Kazuhiro and {Okamura}, Sadanori and {Furusawa}, Hisanori and {Kashikawa}, Nobunari and {Ota}, Kazuaki and {Doi}, Mamoru and {Hamabe}, Masaru and {Kimura}, Masahiko and {Komiyama}, Yutaka and {Miyazaki}, Masayuki and {Miyazaki}, Satoshi and {Nakata}, Fumiaki and {Sekiguchi}, Maki and {Yagi}, Masafumi and {Yasuda}, Naoki},
        title = "{Subaru Deep Survey. V. A Census of Lyman Break Galaxies at z\raisebox{-0.5ex}\textasciitilde=4 and 5 in the Subaru Deep Fields: Photometric Properties}",
      journal = {\apj},
         year = 2004,
        month = aug,
       volume = {611},
       number = {2},
        pages = {660-684},
          doi = {10.1086/422207},
archivePrefix = {arXiv},
       eprint = {astro-ph/0309655},
 primaryClass = {astro-ph},
       adsurl = {https://ui.adsabs.harvard.edu/abs/2004ApJ...611..660O}
}

@ARTICLE{Bahcall_1983,
       author = {{Bahcall}, J.~N. and {Soneira}, R.~M.},
        title = "{The universe at faint magnitudes. I. Models for the Galaxy and the predicted star counts.}",
      journal = {\apjs},
         year = 1980,
        month = sep,
       volume = {44},
        pages = {73-110},
          doi = {10.1086/190685},
       adsurl = {https://ui.adsabs.harvard.edu/abs/1980ApJS...44...73B}
}

@ARTICLE{Davis_1983,
       author = {{Davis}, M. and {Peebles}, P.~J.~E.},
        title = "{A survey of galaxy redshifts. V. The two-point position and velocity correlations.}",
      journal = {\apj},
         year = 1983,
        month = apr,
       volume = {267},
        pages = {465-482},
          doi = {10.1086/160884},
       adsurl = {https://ui.adsabs.harvard.edu/abs/1983ApJ...267..465D}
}

@ARTICLE{Bardeen_1986,
       author = {{Bardeen}, J.~M. and {Bond}, J.~R. and {Kaiser}, N. and {Szalay}, A.~S.},
        title = "{The Statistics of Peaks of Gaussian Random Fields}",
      journal = {\apj},
         year = 1986,
        month = may,
       volume = {304},
        pages = {15},
          doi = {10.1086/164143},
       adsurl = {https://ui.adsabs.harvard.edu/abs/1986ApJ...304...15B}
}

@ARTICLE{Dalmasso_2024,
       author = {{Dalmasso}, Nicol{\`o} and {Leethochawalit}, Nicha and {Trenti}, Michele and {Boyett}, Kristan},
        title = "{Galaxy clustering at cosmic dawn from JWST/NIRCam observations to redshift z 11}",
      journal = {\mnras},
         year = 2024,
        month = sep,
       volume = {533},
       number = {2},
        pages = {2391-2398},
          doi = {10.1093/mnras/stae2006},
archivePrefix = {arXiv},
       eprint = {2402.18052},
 primaryClass = {astro-ph.GA},
       adsurl = {https://ui.adsabs.harvard.edu/abs/2024MNRAS.533.2391D}
}

@ARTICLE{HOD_Ma_Fry_2000,
       author = {{Ma}, Chung-Pei and {Fry}, J.~N.},
        title = "{Deriving the Nonlinear Cosmological Power Spectrum and Bispectrum from Analytic Dark Matter Halo Profiles and Mass Functions}",
      journal = {\apj},
         year = 2000,
        month = nov,
       volume = {543},
       number = {2},
        pages = {503-513},
          doi = {10.1086/317146},
archivePrefix = {arXiv},
       eprint = {astro-ph/0003343},
 primaryClass = {astro-ph},
       adsurl = {https://ui.adsabs.harvard.edu/abs/2000ApJ...543..503M}
}

@ARTICLE{HOD_Peacock_Smith_2000,
       author = {{Peacock}, J.~A. and {Smith}, R.~E.},
        title = "{Halo occupation numbers and galaxy bias}",
      journal = {\mnras},
         year = 2000,
        month = nov,
       volume = {318},
       number = {4},
        pages = {1144-1156},
          doi = {10.1046/j.1365-8711.2000.03779.x},
archivePrefix = {arXiv},
       eprint = {astro-ph/0005010},
 primaryClass = {astro-ph},
       adsurl = {https://ui.adsabs.harvard.edu/abs/2000MNRAS.318.1144P}
}

@ARTICLE{HOD_Seljak_2000,
       author = {{Seljak}, Uro{\v{s}}},
        title = "{Analytic model for galaxy and dark matter clustering}",
      journal = {\mnras},
         year = 2000,
        month = oct,
       volume = {318},
       number = {1},
        pages = {203-213},
          doi = {10.1046/j.1365-8711.2000.03715.x},
archivePrefix = {arXiv},
       eprint = {astro-ph/0001493},
 primaryClass = {astro-ph},
       adsurl = {https://ui.adsabs.harvard.edu/abs/2000MNRAS.318..203S}
}

@ARTICLE{Park_2017,
       author = {{Park}, Jaehong and {Kim}, Han-Seek and {Liu}, Chuanwu and {Trenti}, Michele and {Duffy}, Alan R. and {Geil}, Paul M. and {Mutch}, Simon J. and {Poole}, Gregory B. and {Mesinger}, Andrei and {Wyithe}, J. Stuart B.},
        title = "{Dark-ages reionization and galaxy formation simulation-XI. Clustering and halo masses of high redshift galaxies}",
      journal = {\mnras},
         year = 2017,
        month = dec,
       volume = {472},
       number = {2},
        pages = {1995-2008},
          doi = {10.1093/mnras/stx1884},
archivePrefix = {arXiv},
       eprint = {1703.05419},
 primaryClass = {astro-ph.GA},
       adsurl = {https://ui.adsabs.harvard.edu/abs/2017MNRAS.472.1995P}
}

@ARTICLE{Zheng_2005,
       author = {{Zheng}, Zheng and {Berlind}, Andreas A. and {Weinberg}, David H. and {Benson}, Andrew J. and {Baugh}, Carlton M. and {Cole}, Shaun and {Dav{\'e}}, Romeel and {Frenk}, Carlos S. and {Katz}, Neal and {Lacey}, Cedric G.},
        title = "{Theoretical Models of the Halo Occupation Distribution: Separating Central and Satellite Galaxies}",
      journal = {\apj},
         year = 2005,
        month = nov,
       volume = {633},
       number = {2},
        pages = {791-809},
          doi = {10.1086/466510},
archivePrefix = {arXiv},
       eprint = {astro-ph/0408564},
 primaryClass = {astro-ph},
       adsurl = {https://ui.adsabs.harvard.edu/abs/2005ApJ...633..791Z}
}

@ARTICLE{Garel_2015,
       author = {{Garel}, T. and {Blaizot}, J. and {Guiderdoni}, B. and {Michel-Dansac}, L. and {Hayes}, M. and {Verhamme}, A.},
        title = "{The UV, Lyman {\ensuremath{\alpha}}, and dark matter halo properties of high-redshift galaxies}",
      journal = {\mnras},
         year = 2015,
        month = jun,
       volume = {450},
       number = {2},
        pages = {1279-1294},
          doi = {10.1093/mnras/stv374},
archivePrefix = {arXiv},
       eprint = {1503.06635},
 primaryClass = {astro-ph.GA},
       adsurl = {https://ui.adsabs.harvard.edu/abs/2015MNRAS.450.1279G}
}

@ARTICLE{Bartelmann_2001,
       author = {{Bartelmann}, M. and {Schneider}, P.},
        title = "{Weak gravitational lensing}",
      journal = {\physrep},
         year = 2001,
        month = jan,
       volume = {340},
       number = {4-5},
        pages = {291-472},
          doi = {10.1016/S0370-1573(00)00082-X},
archivePrefix = {arXiv},
       eprint = {astro-ph/9912508},
 primaryClass = {astro-ph},
       adsurl = {https://ui.adsabs.harvard.edu/abs/2001PhR...340..291B}
}

@ARTICLE{Cooray_2002,
       author = {{Cooray}, Asantha and {Sheth}, Ravi},
        title = "{Halo models of large scale structure}",
      journal = {\physrep},
         year = 2002,
        month = dec,
       volume = {372},
       number = {1},
        pages = {1-129},
          doi = {10.1016/S0370-1573(02)00276-4},
archivePrefix = {arXiv},
       eprint = {astro-ph/0206508},
 primaryClass = {astro-ph},
       adsurl = {https://ui.adsabs.harvard.edu/abs/2002PhR...372....1C}
}

@ARTICLE{Correa_2015,
       author = {{Correa}, Camila A. and {Wyithe}, J. Stuart B. and {Schaye}, Joop and {Duffy}, Alan R.},
        title = "{The accretion history of dark matter haloes - III. A physical model for the concentration-mass relation}",
      journal = {\mnras},
         year = 2015,
        month = sep,
       volume = {452},
       number = {2},
        pages = {1217-1232},
          doi = {10.1093/mnras/stv1363},
archivePrefix = {arXiv},
       eprint = {1502.00391},
 primaryClass = {astro-ph.CO},
       adsurl = {https://ui.adsabs.harvard.edu/abs/2015MNRAS.452.1217C}
}


\appendix


\bsp	
\label{lastpage}
\end{document}